\documentclass[preprint,11pt,number]{elsarticle}
\usepackage[utf8]{inputenc}
\usepackage{amsmath}
\usepackage{amssymb}
\usepackage{amsthm}
\usepackage{graphicx}
\usepackage{listings}
\usepackage{appendix}
\usepackage{natbib}
\usepackage{array}
\usepackage{lipsum}
\usepackage{stmaryrd}
\usepackage{longtable}
\usepackage{multirow}
\usepackage{tikz}
\usepackage{caption}
\usepackage{subcaption}
\usepackage{lineno}
\usepackage{algorithm}
\usepackage[noend]{algpseudocode}
\usepackage{mathtools}
\usepackage{fullpage}
\usepackage{indentfirst}
\usepackage{xcolor}
\usepackage{courier} 
\usepackage{float}
\usepackage{hyperref}
\usepackage{caption} 
\usepackage{xcolor}
\usepackage{doi}
\usepackage{booktabs} 
\usepackage{orcidlink}
\newcommand{\abs}[1]{\lvert #1 \rvert}

\usetikzlibrary{shapes.geometric, arrows}
\tikzstyle{startstop} = [rectangle, rounded corners, minimum width=3.5cm, minimum height=0.5cm, text centered, draw=black, fill=red!20, font=\small]
\tikzstyle{process} = [rectangle, minimum width=3.5cm, minimum height=1cm, text centered, draw=black, fill=blue!20, font=\small]
\tikzstyle{data} = [rectangle, rounded corners, minimum width=3.5cm, minimum height=1cm, text centered, draw=black, fill=gray!20, font=\small]
\tikzstyle{arrow} = [thick,->,>=stealth]
\theoremstyle{definition}

\theoremstyle{remark}

\theoremstyle{plain}

\newcommand{\cyr}[1]{\textcolor{black}{#1}}
\begin{document}
\begin{frontmatter}

\title{Stochastic Coefficient of Variation: Assessing the Variability and Forecastability of Solar Irradiance} 

\author[label1]{Cyril Voyant\corref{cor1}\orcidlink{0000-0003-0242-7377}}
\ead{cyril.voyant@mines-psl.eu}
\cortext[cor1]{Corresponding Author}
\author[label1]{Alan Julien}
\ead{alan.julien@mines-psl.eu}
\author[label2,label4]{Milan Despotovic\orcidlink{0000-0003-3144-5945}}
\ead{mdespotovic@kg.ac.rs} 
\author[label4]{Gilles Notton\orcidlink{0000-0002-6267-9632}}
\ead{notton_g@univ-corse.fr}
\author[label4]{Luis Antonio Garcia-Gutierrez\orcidlink{0000-0002-3480-1784}}
\ead{garcia_gl@univ-corse.fr} 
\author[label4,label5]{Claudio Francesco Nicolosi\orcidlink{0009-0001-2962-6643}}
\ead{claudio.nicolosi@phd.unict.it}
\author[label1]{Philippe Blanc\orcidlink{0000-0002-6345-0004}}
\ead{philippe.blanc@mines-psl.eu}
\author[label6]{Jamie Bright\orcidlink{0000-0002-9465-3453}}
\ead{jamiebright1@gmail.com} 

\address[label1]{Mines Paris, \texttt{PSL} University, Centre for observation, impacts, energy (\texttt{O.I.E.}), 06904 Sophia Antipolis, France}
\address[label2]{Faculty of Engineering, University of Kragujevac, 6 Sestre Janjic, Kragujevac, Serbia}
\address[label4]{\texttt{SPE} Laboratory, \texttt{UMR CNRS} 6134, University of Corsica Pasquale Paoli, Ajaccio, France}
\address[label5]{Department of Electrical, Electronic and Computer Engineering (\texttt{DIEEI}), University of Catania, Viale Andrea Doria n.6, 95125 Catania, Italy}
\address[label6]{Distribution System Operator, \texttt{UK Power Networks}, London, United Kingdom}

\date{\today}
\journal{Renewable Energy}
\begin{abstract}
This work presents a robust framework for quantifying solar irradiance variability and forecastability through the Stochastic Coefficient of Variation ($\mathtt{sCV}$) and the Forecastability ($\mathtt{F}$). Traditional metrics, such as the standard deviation, fail to isolate stochastic fluctuations from deterministic trends in solar irradiance. By considering clear-sky irradiance as a dynamic upper bound of measurement, $\mathtt{sCV}$ provides a normalized, dimensionless measure of variability that theoretically ranges from 0 to 1. $\mathtt{F}$ extends $\mathtt{sCV}$ by integrating temporal dependencies via maximum autocorrelation, thus linking $\mathtt{sCV}$ with $\mathtt{F}$. The proposed methodology is validated using synthetic cyclostationary time series and experimental data from 68 meteorological stations (in Spain). Our comparative analyses demonstrate that $\mathtt{sCV}$ and $\mathtt{F}$ proficiently encapsulate multi-scale fluctuations, while addressing significant limitations inherent in traditional metrics. This comprehensive framework enables a refined quantification of solar forecast uncertainty, supporting improved decision-making in flexibility procurement and operational strategies. By assessing variability and forecastability across multiple time scales, it enhances real-time monitoring capabilities and informs adaptive energy management approaches, such as dynamic outage management and risk-adjusted capacity allocation.
\end{abstract}

\begin{graphicalabstract}
\vspace{2em}
\begin{tikzpicture}[node distance=2.2cm]
\tikzstyle{startstop} = [rectangle, rounded corners, minimum width=2.5cm, minimum height=0.5cm, text centered, draw=black, fill=red!20, font=\small]
\tikzstyle{process} = [rectangle, minimum width=3.5cm, minimum height=1cm, text centered, draw=black, fill=blue!20, font=\small]
\tikzstyle{data} = [rectangle, rounded corners, minimum width=3.5cm, minimum height=1cm, text centered, draw=black, fill=gray!20, font=\small]
\tikzstyle{arrow} = [thick,->,>=stealth]
\node (input1) [data] {Input: $\mathtt{I}(t)$};
\node[anchor=north west, xshift=-2.3cm, yshift=-1cm] {\includegraphics[width=4.5cm]{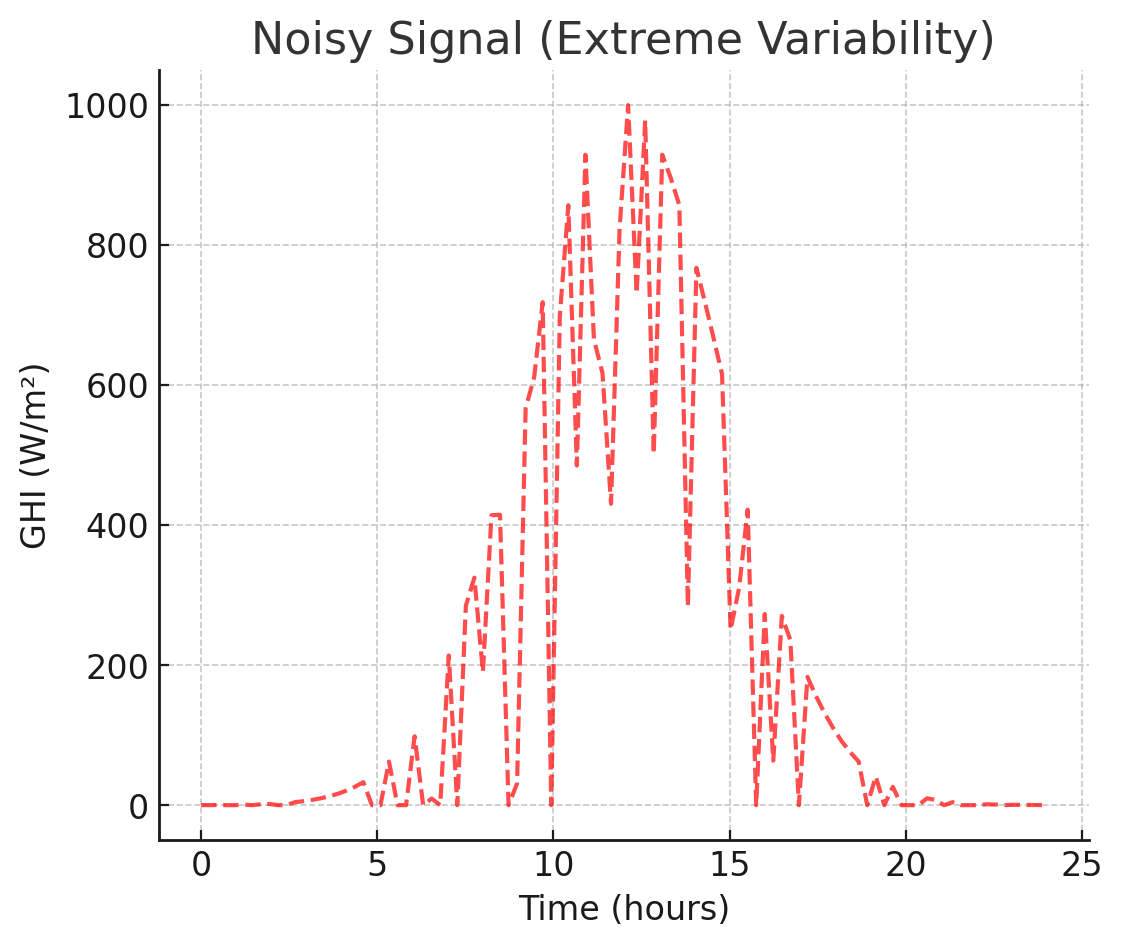}};
\node[anchor=north west, xshift=10.0cm, yshift=-1cm] {\includegraphics[width=4.5cm]{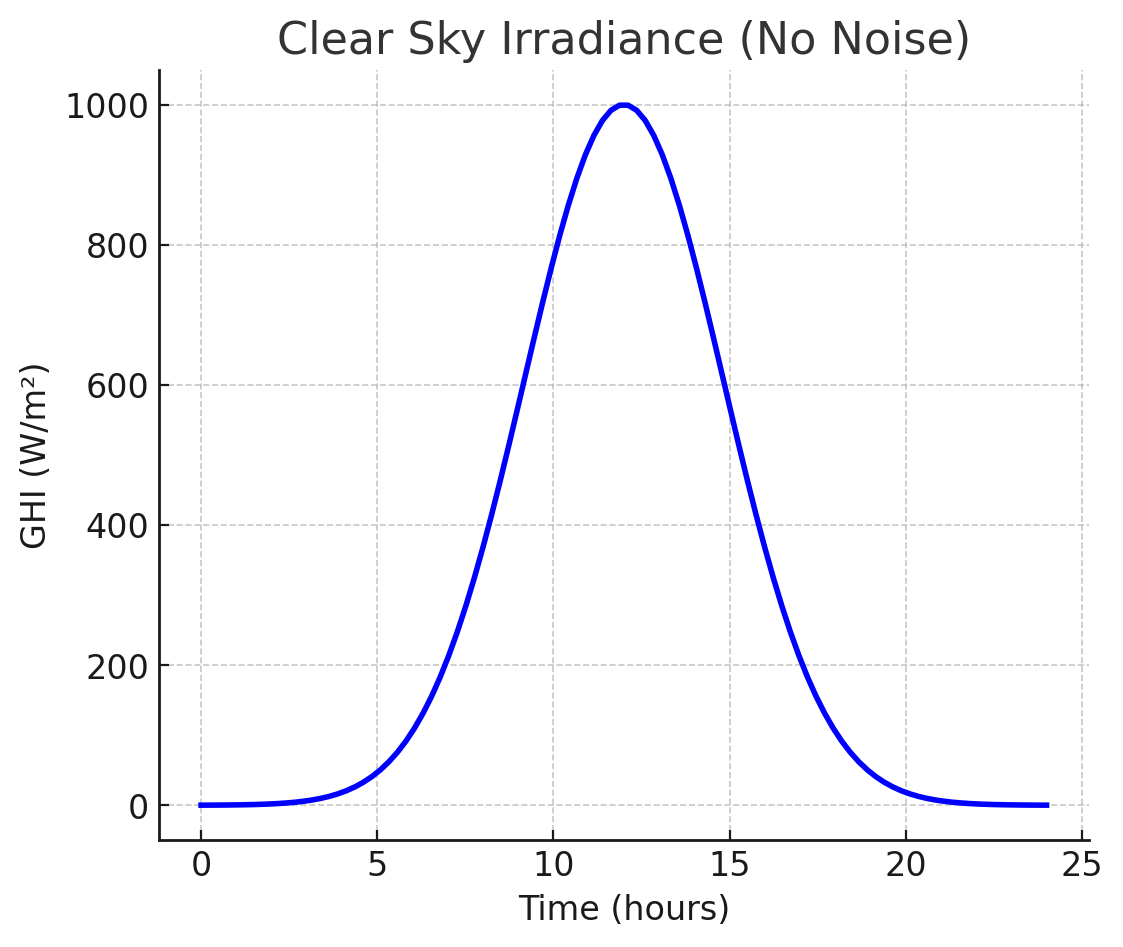}};
\node (deviation) [process, right of=input1, xshift=4cm] {Compute Deviation:  $\mathtt{I}(t) - \mathtt{I}_{\mathtt{clr}}(t)$};
\node (input2) [data, right of=deviation, xshift=4cm] {Input: $\mathtt{I}_{\mathtt{clr}}(t)$};
\node (variability) [process, below of=deviation, yshift=0.5cm] {Compute Variability:  $\sigma(\mathtt{I}) = \sqrt{\mathbb{E}[(\mathtt{I} - \mathtt{I}_{\mathtt{clr}})^2]}$};
\node (scv) [process, below of=variability, yshift=0.5cm] {Normalize:  $\mathtt{sCV} = \frac{\sqrt{2}}{\abs{d-1}  \sqrt{\sigma^2_{\mathtt{clr}}+\mu_{\mathtt{clr}}^2}}  \sigma(\mathtt{I})$};
\node (forecastability) [process, below of=scv, yshift=0.5cm] {Compute Forecastability:  $\mathtt{F} = (1 - \mathtt{sCV}) +\rho_{max} \mathtt{sCV}$};
\node (applications) [startstop, below of=forecastability, yshift=0.5cm] {Applications:  Variability Analysis, Solar Integration, Forecast Validation};
\draw [arrow] (input1.east) -- (deviation.west);
\draw [arrow] (input2.west) -- (deviation.east);
\draw [arrow] (deviation.south) -- (variability.north);
\draw [arrow] (variability.south) -- (scv.north);
\draw [arrow] (scv.south) -- (forecastability.north);
\draw [arrow] (forecastability.south) -- (applications.north);
\end{tikzpicture}
\centering
\fbox{\parbox{\textwidth}{\centering\textbf{Framework for Quantifying Solar Variability ($\mathtt{sCV})$ and Forecastability ($\mathtt{F}$)}}}
\end{graphicalabstract}

\begin{highlights}
    \item \cyr{New metric quantifying solar variability via Stochastic Coefficient of Variation ($\mathtt{sCV}$)
    \item Clear-sky irradiance ($\mathtt{I}_{\mathtt{clr}}$) as an upper bound for fluctuations
    \item Normalization with $\mathtt{sCV}$, bounded between 0 and 1, suitable for non-stationary signals
    \item Linking variability and predictability through forecastability ($\mathtt{F}$)
    \item Validation on synthetic data and observations from 68 stations}
\end{highlights}


\end{frontmatter}
\section{Introduction}
\label{intro}
In the dynamic world of renewable energies, understanding and quantifying variability is not just an academic exercise: it is a key to unlocking reliable and cost-effective solar power. For energy professionals, accurately characterizing the fluctuating nature of solar irradiance is important, whether for optimizing grid integration or minimizing energy storage costs \cite{Perez2010, Voyant2017}. This article explores the essential role of variability estimation in solar energy time series, integrating established and emerging methodologies to provide actionable information in the field. Despite significant advances in solar forecasting techniques, there remains a critical gap in capturing short-term and multi-scale fluctuations \cite{kopp2016magnitudes,blaga2018quantifiers} with sufficient robustness for operational applications. This gap becomes evident when rapid changes in solar irradiance, often driven by cloud transients or local weather events \cite{lauret2016characterization,aguera2024meteorological}, necessitate real-time adjustments. Moreover, long-term variations (associated with seasonal or climatological shifts) must also be accurately characterized to inform capacity planning and investment decisions \cite{Lauret2016}. In particular, anticipating day-ahead variability on a site-specific basis is crucial for optimizing the procurement strategy in flexibility markets. A refined knowledge of local forecast uncertainty enables operators to adjust their conservatism levels when securing flexibility resources. Furthermore, such predictive capabilities underpin advanced operational strategies such as dynamic outage management, where forecast-driven decision-making allows for the temporary restoration of generating capacity during planned outages. For sites exhibiting high forecast predictability, additional capacity can be safely released, whereas more conservative procurement strategies must be adopted for locations with greater uncertainty.

\subsection{\cyr{Text Mining Context: The Research Gaps}}
\cyr{To establish the scientific relevance of our research on solar irradiance variability and forecastability, a targeted text mining and bibliometric analysis of the literature published between 2020 and 2024 was conducted. This approach quantifies the current state of knowledge and highlights the research gap that motivates the introduction of our new metrics.} Based on 624 publications retrieved from \texttt{Scopus} and enriched via \texttt{CrossRef} and \texttt{OpenAlex}, the approach combined keyword-based filtering with semantic scoring using transformer-based embeddings (\texttt{MiniLM}), ensuring both relevance and breadth. 
The analysis focused on four core concepts: \textit{variability}, \textit{volatility}, \textit{forecastability}, and \textit{predictability}. From 2020 to 2024, the use of these terms increased steadily, reaching its peak in 2024 with over 120 occurrences. Variability dominates by far, appearing in over 70\% of relevant records. Volatility follows with steady growth, while the latter two terms remain niche, indicating limited exploration of deeper predictability metrics in the current literature. Publication formats are led by journal articles, which represent 69\% of the corpus, followed by conference papers (19\%). The remainder consists of reviews, book chapters, and technical notes. This distribution confirms the academic maturity of the field, while the presence of conference proceedings reflects its ongoing dynamism and practical orientation. In terms of geographic data focus, 39\% of studies analyzed datasets from Asia, followed by 17\% focused on Europe and 10\% on North America. Africa accounted for 7\%, while South America and Oceania remained marginal (2\% each). Notably, nearly a quarter of the studies (23\%) concentrated on specific data rather than broader regional contexts, such as national labs or agency-specific datasets.
Time series modeling stands out as the primary methodological choice in more than three-quarters of identified cases, reflecting the operational focus on historical trend-based forecasting. Other techniques such as numerical weather prediction or AI-driven hybrid methods are mentioned but remain marginal. This suggests a field still anchored in deterministic or statistical baselines rather than probabilistic or physics-informed modeling. 
Geographically, Asia leads both in volume and impact, with China alone contributing over 70\% of publications and citations. India and the United States follow distantly (15 and 16\% respectively). The most cited institutions over the period include \texttt{Harbin Institute of Technology} (45\% of citations among the top three), \texttt{Clean Power Research} (30\%), and the \texttt{University of Aveiro} (25\%), reflecting a concentration of influence among a few technically oriented and internationally visible research centers.
Publication venues are dominated by \texttt{Elsevier} and \texttt{MDPI}, which account more than 50\% \cyr{of} all publications. Key journals include \textit{Energies}, \texttt{Renewable Energy}, and \texttt{Sustainability}, reflecting a strong presence of energy-focused outlets within a fragmented journal landscape.
 Around 43\% of the publications were open access, signaling a gradual shift toward greater accessibility in the field. Among the most cited articles, a comprehensive review of solar forecasting methods stands out with 166 citations \cite{YANG2022112348}. Journal dispersion is high, with few titles repeatedly publishing on the topic, indicating a cross-disciplinary reach spanning energy, meteorology, and environmental systems. 
On the author side, output is dispersed. A few key individuals stand out in terms of citations, including Dazhi Yang (\texttt{Harbin Institute of Technology}, China), Jing Huang (\texttt{Clean Power Research}, USA), and D. Van Der Meer (\texttt{Mines-PSL University}, France), whose institutional affiliations correspond to those at the time of publication. While these figures emerge prominently, the majority of contributions remain single-entry efforts, underscoring the field wide distribution and the absence of a consolidated author core. Highly cited articles focus on variability but also on short-term solar forecasting using advanced multi-step or hybrid methods, underlining the practical relevance for grid integration and planning. 
Despite the scale and semantic refinement of this analysis, significant limitations remain. Text mining, even when enhanced with Natural Language Processing (\textit{i.e.} \texttt{NLP}) techniques, cannot fully grasp the nuance or depth of a scientific contribution. Many publications reference variability only superficially (often as a contextual keyword) without treating it as a core subject of investigation. In most cases, no new metric is introduced, and variability is neither formally defined nor methodologically problematized. This highlights a conceptual gap in the literature, where variability is frequently invoked but rarely interrogated. As a result, this type of metadata-based screening (focused on titles, abstracts, and keywords) must be seen as a preliminary diagnostic rather than a substitute for deep, qualitative analysis. It is a powerful tool for mapping the field’s surface, but it cannot replace critical engagement with the content and structure of the studies themselves. \cyr{Finally, this text mining approach shows the limitations of traditional metrics (\textit{e.g.}) and deterministic clear-sky models, which often fail to separate stochastic variability from deterministic trends and do not directly quantify predictability. To address these gaps, new metrics to normalize variability relative to a dynamic upper bound, and to capture the temporal predictability of residual fluctuations will be proposed. Together, these new metrics will provide robust and physically meaningful assessments that remain consistent across different time scales and sites.}

\subsection{Variability in the Solar Energy Context}
Solar irradiance variability, driven by daily, annual and stochastic atmospheric effects \cite{Lauret2016}, impacts system reliability, forecasting, and grid integration \cite{LAVE2012, Perez2011}. Rapid cloud-induced fluctuations cause sudden power ramps, straining grid balancing and requiring reserves or storage. Addressing these challenges requires a multi-scale approach: some methods support long-term planning and $\mathtt{PV}$ site selection, while others focus on real-time monitoring for forecasting and volatility management. 

The first category of metrics captures long-term trends and site-specific characteristics using aggregated data and statistical references like mean, variance or covariance. A widely used measure is the coefficient of variation ($\mathtt{CV}$), defined as the ratio of the standard deviation to the mean irradiance (Global Solar Irradiance, \textit{i.e.} $\mathtt{I}$) over a given period, $\mathtt{CV} = \sigma_{\mathtt{I}} / \mu_{\mathtt{I}}$. Another common metric is the nominal variability $\sigma(\Delta \mathtt{k_c})$, derived from the clear-sky index, where $\Delta \mathtt{k_c} = \mathtt{k_c}(t) - \mathtt{k_c}(t-1)$, with $\mathtt{k_c}$ representing the ratio of measured to clear-sky irradiance \cite{solar2040026}. 
A detailed derivation and its connection to $\mathtt{RMSE}$ in smart persistence prediction can be found in \ref{perez}. These metrics help classify sites based on long-term variability, aiding grid planning and energy storage design \cite{Perez2011,Voyant2015}.  
\cite{stein2012variability} introduced the Variability Index ($\mathtt{VI}$), which quantifies irradiance variability as the ratio of the measured irradiance curve length to the clear-sky curve length, adjusted for measurement intervals to ensure consistency. The Monte Carlo-based forecastability metric from \cite{Voyant2021} integrates clear-sky models with $\mathtt{RMSE}$ to quantify predictability. Similarly, \cite{Perez2018} developed spatio-temporal smoothing models to mitigate anomalies across regions, improving large-scale variability assessments. Statistical methods, such as random forests, have also been applied to model daily variability from meteorological data \cite{Huang2014}.  
While these methods provide a valuable framework for strategic planning, they have limitations. Their reliance on aggregated data (\textit{e.g.}, daily or monthly averages) can obscure high-frequency fluctuations crucial for real-time operations. Additionally, inaccuracies in clear-sky models or timestamp misalignments may introduce biases, reducing the reliability of variability estimates \cite{Lauret2016}.

The second category of metrics focuses on high-frequency dynamics, providing real-time insights into irradiance variability and making them essential for system operations. Metrics such as the mean absolute log-return ($\mathtt{MALR}$) capture short-term fluctuations and are defined as $\mathtt{MALR} = \frac{1}{N} \sum_{t=1}^{N-1} \left| \log \left( \frac{\mathtt{I}(t)}{\mathtt{I}(t-1)} \right) \right|$, where $\mathtt{I}(t)$ is the irradiance at time $t$. Other measures, such as the Daily Aggregate Ramp Rate ($\mathtt{DARR}$), provide a cumulative assessment of power changes over time, offering practical insights for grid management \cite{VanHaaren2014}.  
Another ramp rate-based metric is the Variability Score from Ramp Rate Distribution ($\mathtt{VSRR}_{\text{dist}}$), introduced by Lave \textit{et al.} \cite{LAVE2015}, which quantifies solar variability by combining the maximum ramp rate magnitude with its probability of occurrence, scaled for interpretability. Similarly, binary sky-state variables can be used to develop metrics such as the Daily Aggregate Sunshine State Number ($\mathtt{DASSN}$) \cite{blaga2018quantifiers}, which measures changes in sun coverage by tracking transitions between clear and covered sky states based on irradiance thresholds.  
These methods are particularly valuable in high-variability environments requiring rapid adjustments. For instance, \cite{McCandless2015} demonstrated that incorporating temporal and spatial standard deviations into forecasting models improves real-time grid balancing. Likewise, \cite{Kreuwel2020} analyzed ramp rates and peak fluctuations, emphasizing their operational relevance under mixed-cloud conditions. By focusing on short-term dynamics, these metrics enable adaptive responses to fluctuating solar resources.  
Despite their advantages, these methods (rooted in local dynamics) lack the broader context provided by first kind of approaches (more global). Metrics such as $\mathtt{MAL}$ and $\mathtt{DARR}$ effectively detect immediate variability but are less suited for long-term planning or regional assessments. Their dependence on sequential data also limits their ability to capture variability induced by external factors like regional weather patterns. In this sense, probabilistic models such as those by \cite{ALONSOSUAREZ2020554} bridge these scales by integrating satellite data with ground-based measurements, addressing both short-term fluctuations and long-term trends. Numerous other studies have proposed methods for estimating renewable variability \cite{en17184549}, stability or volatility; however, these approaches have not been widely adopted in the literature \cite{stein2012variability}. \cyr{This is likely due to their inherent complexity \cite{2010_PAULESCU} or underlying theoretical limitations \cite{wind}, which have hindered their practical implementation \cite{blaga2018quantifiers} and broader recognition \cite{LAVE2015,LAVE2012}. A new promising class of approaches relies on entropy-based metrics such as Weighted Permutation Entropy ($\mathtt{WPE}$), which has been proposed as a predictor of site-level forecast error \cite{Karimi2023}. While these methods are conceptually appealing, their practical utility remains limited by several factors. Notably, $\mathtt{WPE}$ values often exhibit a narrow dynamic range across different sites, which reduces their sensitivity to local variability. Furthermore, entropy-based metrics typically require careful hyperparameter tuning and may lack physical interpretability. As a result, although $\mathtt{WPE}$-based models can effectively capture certain aspects of forecast error, they may be less intuitive and actionable, particularly in early deployment contexts where interpretability and robustness are critical.
}

\section{Context \& Motivation}  
\label{sec:context}
The integration of solar photovoltaics reduces confidence in net load forecasting, impacting flexibility markets and balancing services that rely on short-term predictions. Increased uncertainty may necessitate more conservative forecasts, leading to higher balancing costs. Existing methods for variability estimation face mathematical and operational limitations, restricting their effectiveness and applicability.
One key issue is the reliance on the clear-sky index ($\mathtt{K_c}$). While widely used, $\mathtt{K_c}$ is highly sensitive to misalignments between measured data and predictions. Timestamp errors (due to sensor desynchronization or model parameterization) artificially inflate variability metrics, making observed fluctuations reflect misalignment rather than true stochastic variations, thus compromising physical interpretability.
Another major limitation of $\mathtt{K_c}$ is its instability under low-irradiance conditions (sunrise or sunset). The clear-sky model then produces very low values, exaggerating minor measurement errors into large $\mathtt{K_c}$ peaks, creating spurious variability unrelated to atmospheric dynamics. Moreover, since $\mathtt{K_c}$ is undefined at night, filtering nighttime data relies on subjective thresholds (\textit{e.g.}, solar elevation of 5° or 10°), affecting variability estimates and leading to inconsistencies across studies. Such subjective filtering undermines rigorous and reproducible analysis.
From a mathematical perspective, the multiplicative nature of $\mathtt{K_c}$ distorts the assessment of variability, especially under previously mentioned unstable conditions. Cloud transients induce exaggerated fluctuations in $\mathtt{K_c}$, even when absolute irradiance variations are small, complicating analysis and reducing its operational usefulness. Moreover, $\mathtt{K_c}$ assumes that clear-sky models perfectly capture the deterministic components of irradiance, which is rarely the case due to local atmospheric uncertainties \cite{SUN2021110087,SUN2019550}.
Another issue is the lack of bounds for certain variability metrics. For instance, metrics based on the coefficient of variation ($\mathtt{CV}$) can become infinite when the mean irradiance approaches zero, such as during cloudy periods or low-irradiance conditions. This absence of limits reduces the physical relevance of the metric and complicates comparisons across climates and timescales.
Given these limitations, improved variability metrics are needed to overcome existing shortcomings. Metrics that do not rely on $\mathtt{K_c}$ exist, but for now, they are complex and not widely used or studied. New metrics should avoid multiplicative $\mathtt{K_c}$, remain bounded under all conditions, and capture stochastic fluctuations rather than artifacts from timestamp alignment or data filtering. 
Additionally, they must ensure consistency and reproducibility across datasets without subjective thresholds. \cyr{Addressing these challenges will enhance variability, predictability, and forecastability metrics, ultimately improving forecasting, system design, and grid integration for more reliable and efficient solar energy systems. \ref{properties2} provides a list of desirable properties for such estimators. Traditional metrics, such as the standard deviation and the coefficient of variation ($\mathtt{CV}$), offer valuable insights into variability but remain unbounded. This unbounded nature can cause these metrics to lose interpretability when forecast initial conditions (such as time step and site-specific factors) are not well controlled. Moreover, these metrics do not adequately separate stochastic fluctuations from the deterministic clear-sky trend. To overcome these limitations, we introduce physically grounded and bounded metrics, namely, the stochastic coefficient of variation ($\mathtt{sCV}$) and the forecastability ($\mathtt{F}$), that ensure robust and consistent quantification of solar variability and forecastability across sites and temporal resolutions.}

\section{Objectives}
\label{sec:objectives}
The first objective of this study is to develop and evaluate robust metrics for quantifying solar irradiance variability and forecastability that address the limitations of existing methods. Current approaches often fail to accurately represent variability due to their reliance on $\mathtt{k_c}$, sensitivity to timestamp alignment errors, and artifacts arising from multiplicative definitions and subjective data filtering. These shortcomings hinder the ability to capture true stochastic fluctuations and misrepresent the physical dynamics of irradiance, leading to inaccuracies in forecasting.
In response, this work aims to establish metrics that are bounded, physically interpretable, and scalable across temporal and spatial resolutions. The proposed metrics will reliably distinguish stochastic fluctuations from deterministic patterns and noise, ensuring robust applicability under diverse climatic and operational conditions. Additionally, they will avoid artifacts associated with multiplicative definitions and timestamp misalignments. These metrics are intended to integrate seamlessly into forecasting frameworks, either by serving as input features to enhance model responsiveness or by providing criteria for selecting predictive models based on observed variability levels. Finally, computational efficiency will also be a key consideration, ensuring that the metrics remain adaptable to both high-frequency data and aggregated temporal resolutions.
In this paper, predictability refers to the ability to construct a forecasting model, closely linked to the Lyapunov horizon, which defines the time limit beyond which predictions become unreliable due to system sensitivity. Between fully predictable and unpredictable states, an infinite range of intermediate conditions exists. This gradation is quantified using the forecastability parameter, which measures how well a system retains predictive power over time.

\section{Methods}
\label{sec:methodology}
The variability of Global Horizontal Irradiance ($\mathtt{I}$) is influenced by a deterministic trend, represented by the clear-sky irradiance ($\mathtt{I}_{\mathtt{clr}}$), which serves as an upper bound \cite{amt-6-2403-2013}. Traditional metrics like the standard deviation or coefficient of variation \cite{variation} fail to distinguish between this deterministic component and the random fluctuations in $\mathtt{I}$. $\mathtt{I}_{\mathtt{clr}}$ refers to the theoretical irradiance under cloud-free conditions ($\sim$ clear-sky), accounting for molecular, aerosols and water vapor scattering. Despite the absence of clouds, $\mathtt{I}_{\mathtt{clr}}$ exhibits minor variability due to atmospheric fluctuations and modeling uncertainties, distinguishing it from a perfectly stable reference.  \cyr{However, the minute-scale variability of atmospheric aerosols is typically smoothed out in these models because they rely on averaged climatological data. Cases of overirradiance, where $\mathtt{I} > \mathtt{I}_{\mathtt{clr}}$ due to cloud-edge effects, should be excluded from clear-sky conditions, even though they can occur in practice \cite{SMITH201710}. This assumption is made to develop the theoretical framework, as these events represent less than 2–5\,\% of cases for time steps greater than 5\,minutes. It should be noted that the true share of pure overirradiance is likely even lower, since this percentage also includes possible biases in the clear-sky model estimation itself.}

\subsection{New Metric for Variability Evaluation}
\begin{figure}
        \centering
        \includegraphics[width=0.6\textwidth]{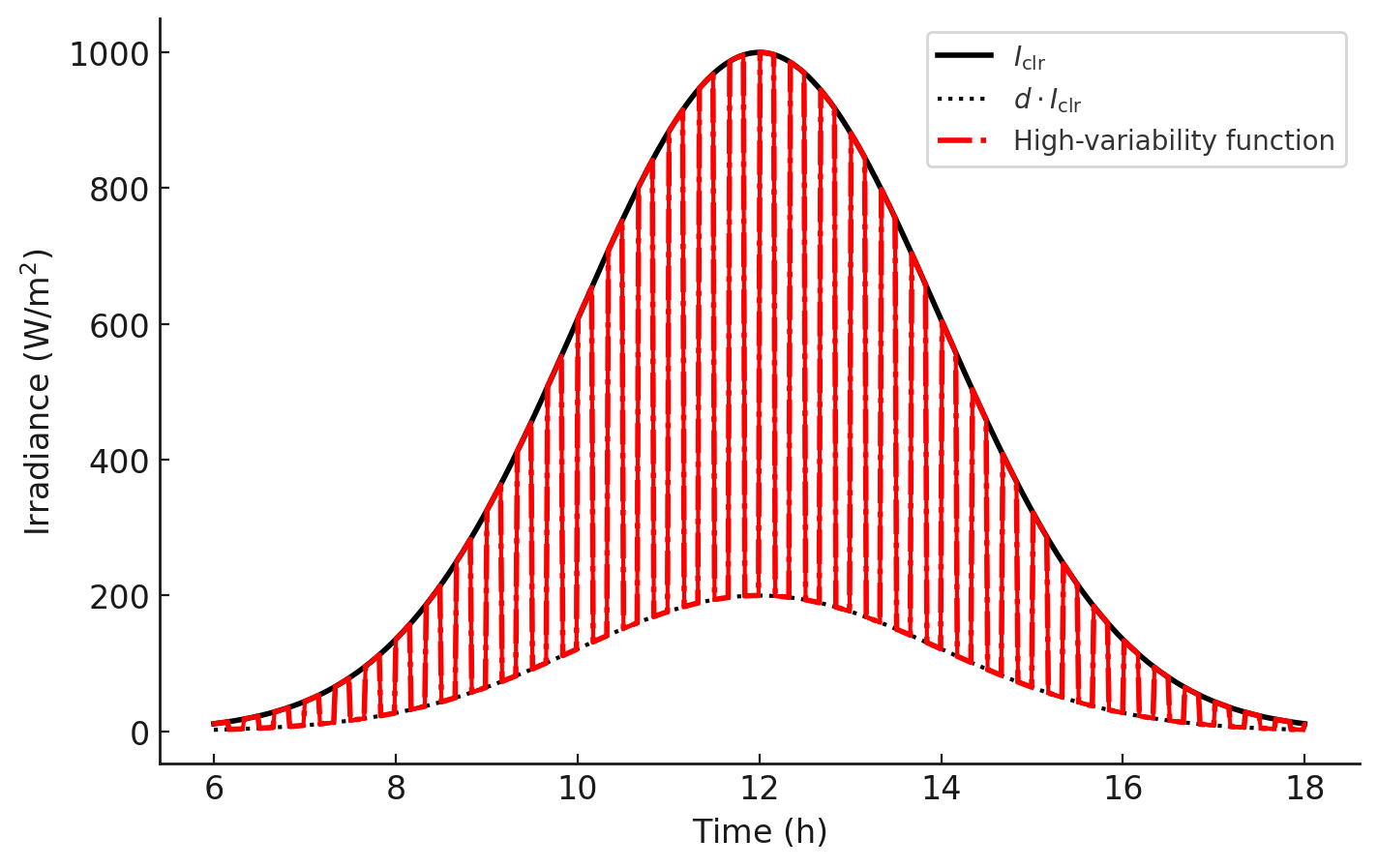}
        \caption{ \cyr{Illustration of the high-variability case. The solid black curve represents the clear-sky irradiance ($\mathtt{I}_{\mathtt{clr}}$), the dotted black line represents the lower irradiance bound ($d \times \mathtt{I}_{\mathtt{clr}}$), and the red dashed area represents a high-variability function $\mathtt{I}(t)$ alternating between clear-sky and diffuse conditions.}}
        \label{worst}
\end{figure}

Let $\mathtt{I}(t)$ be a time series bounded above by the clear-sky irradiance $\mathtt{I}_{\mathtt{clr}}(t)$. To quantify variability relative to this dynamic upper bound, the stochastic coefficient of variation \cyr{is defined by}  $\mathtt{sCV} = \alpha \cdot \sigma(\mathtt{I}),$ with $\alpha$ the normalization constant (designed to scale $\mathtt{sCV}$ between 0 and 1), and
where $\sigma(\mathtt{I})$ is the standard deviation of $\mathtt{I}(t)$ relative to $\mathtt{I}_{\mathtt{clr}}(t)$ defined in a non-standard framework where the reference value is not the mean but a dynamic upper bound or trend:
\begin{equation}
\sigma(\mathtt{I}) = \sqrt{\mathbb{E}\left[(\mathtt{I}(t) - \mathtt{I}_{\mathtt{clr}}(t))^2\right]}.
\end{equation}
When $\mathtt{I} = \mathtt{I}_{\mathtt{clr}}$, the stochastic coefficient of variation ($\mathtt{sCV}$) must be set to $0$, as there are no stochastic fluctuations. Conversely, to determine the normalization constant $\alpha$, which ensures that $\mathtt{sCV}$ remains within the range $[0,1]$ through an appropriate scaling or transformation, the ``worst-case'' scenario \cyr{is considered}. In this case, the maximum variability is assumed ($\textit{e.g.,} $ equal to 1 in Figure~\ref{worst}). This corresponds to a situation where $\mathtt{I}(t)$ alternates between clear-sky conditions $\mathtt{I}_{\mathtt{clr}}(t)$ and cloudy conditions $d \times \mathtt{I}_{\mathtt{clr}}(t)$, with equal probabilities ($50\%$). The value of $d$ represents the diffuse fraction of irradiance since, in theory, $\mathtt{I}(t)$ cannot fall below this value under cloudy conditions. \cyr{In other words, $d$ is used to define the relative lower bound of irradiance, corresponding to the diffuse fraction that remains even under total cloud cover (see Equation~\ref{eq:1}). It is important to note that $d$ itself is a dimensionless factor, not an irradiance value}. Thus, in the considered extreme scenario, $\mathtt{I}$ cannot drop below $d \times \mathtt{I}_{\mathtt{clr}}(t)$. In this extreme case, $\mathtt{sCV}$ is set to its maximum value, $\mathtt{sCV} = 1$. In practice, the measured $\mathtt{I}(t)$ can occasionally drop below $d \times \mathtt{I}_{\mathtt{clr}}$ due to extreme atmospheric conditions; however, such occurrences are rare and thus neglected in this study.
\cyr{
\begin{equation}
        \mathtt{I}(t) - \mathtt{I}_{\mathtt{clr}}(t) =
        \begin{cases}
        0, & \text{if } \mathtt{I}(t) = \mathtt{I}_{\mathtt{clr}}(t), \\
        (d - 1)  \mathtt{I}_{\mathtt{clr}}(t), & \text{if } \mathtt{I}(t) = d \times \mathtt{I}_{\mathtt{clr}}(t).
        \end{cases}
        \label{eq:1}
\end{equation}
$\mathtt{sCV}$ is, in this fact, given by:
\begin{equation}
\mathtt{sCV} = \alpha \sqrt{\frac{1}{n} \sum_{i=1}^n \left(\mathtt{I}(t_i) - \mathtt{I}_{\mathtt{clr}}(t_i)\right)^2}.
\end{equation}
}
It's important to note that the expectation operator ($\mathbb{E}$) used in the initial definition of $\sigma(\mathtt{I})$ is here approximated by the sample average ($\frac{1}{n} \sum$) \cyr{when one is working with discrete time series} data. This approximation is common and justified when dealing with experimental data. To compute this explicitly for the alternating case, two possible states in the dataset \cyr{are considered:
\begin{enumerate}
\item When $\mathtt{I}(t_i) = \mathtt{I}_{\mathtt{clr}}(t_i)$, the deviation is 
$\left(\mathtt{I}(t_i) - \mathtt{I}_{\mathtt{clr}}(t_i)\right)^2 = 0;$
\item When $\mathtt{I}(t_i) = d \times \mathtt{I}_{\mathtt{clr}}(t_i)$, the deviation is 
$\left(\mathtt{I}(t_i) - \mathtt{I}_{\mathtt{clr}}(t_i)\right)^2 = (d - 1)^2 \times \left(\mathtt{I}_{\mathtt{clr}}(t_i)\right)^2$.
\end{enumerate}
}
Both states occur with equal probability ($n_1 = n_2 = n / 2$). The sum of squared deviations over $n$ events can thus be written as:
    \begin{equation}
    \sum_{i=1}^n \left(\mathtt{I}(t_i) - \mathtt{I}_{\mathtt{clr}}(t_i)\right)^2 = \sum_{i \in I_0} 0 + \sum_{i \in I_d} (d - 1)^2 \left(\mathtt{I}_{\mathtt{clr}}(t_i)\right)^2,
    \end{equation}
where $I_0$ and $I_d$ denote the sets of indices corresponding to each case.
Since the second sum involves $n/2$ terms, ($|I_0| = |I_d| = n/2$), \cyr{it follows that}:
\begin{equation}
    \frac{1}{n} \sum_{i=1}^n \left(\mathtt{I}(t_i) - \mathtt{I}_{\mathtt{clr}}(t_i)\right)^2 = \frac{1}{n} (d - 1)^2 \sum_{i=1}^{n/2} (\mathtt{I}_{\mathtt{clr}}(t_{k_i}))^2    ,
\end{equation}
where $t_{k_i}$ represents the timestamps corresponding to the indices in the subset $ I_d$.
According the König-Huygens theorem: $\mathtt{Var}(\mathtt{I}_{\mathtt{clr}})=\sigma^2_{\mathtt{clr}}=\mathbb{E}[\mathtt{I}_{\mathtt{clr}}^2]-\mathbb{E}[\mathtt{I}_{\mathtt{clr}}]^2$ with $\mathtt{Var}$ the conventional variance. \cyr{Considering the mean value of $\mathtt{I}_{\mathtt{clr}}(t_i)$ denoted by $\mu_{\mathtt{clr}}$, it folows}:
\begin{equation}
\frac{1}{n} \sum_{i=1}^n \left(\mathtt{I}(t_i) - \mathtt{I}_{\mathtt{clr}}(t_i)\right)^2 = \frac{(d - 1)^2}{2}  (\sigma^2_{\mathtt{clr}}+\mu_{\mathtt{clr}}^2)
\end{equation}
\cyr{It is important to note that this derivation assumes that the mean and variance of the subset of $I_{clr}$ values are representative of the entire period. This approximation is justified by the periodic alternation of clear-sky and cloudy conditions, along with the approximate symmetry of the clear-sky irradiance curve around solar noon. Such assumptions ensure that any deviations introduced by using only half of the data remain negligible in the long-term analysis. This assumption was validated using clear-sky irradiance data for Sophia Antipolis (France, Lat.\ \(43.614^\circ\)N, Lon.\ \(7.071^\circ\)E). For both hourly and 1-minute data, the difference in the daily mean irradiance resulting from the removal of every other data point was found to be below 1\%, confirming that this approximation introduces negligible bias in long-term variability assessments. It should be noted that for time steps larger than 1h, this approximation may no longer be valid due to the reduced temporal resolution}. Thus, substituting into the expression for $\mathtt{sCV}$, \cyr{it follows}:
\begin{equation}
\mathtt{sCV} = \alpha \sqrt{\frac{1}{2} (d - 1)^2 (\sigma^2_{\mathtt{clr}}+\mu_{\mathtt{clr}}^2)}= \frac{1}{\sqrt{2}} \alpha \abs{d - 1} \sqrt{(\sigma^2_{\mathtt{clr}}+\mu_{\mathtt{clr}}^2)}.
\end{equation}
To normalize $\mathtt{sCV}$ such that $\mathtt{sCV} = 1$ in this alternating scenario, the normalization condition \cyr{is imposed with}:
\begin{equation}
\frac{1}{\sqrt{2}} \alpha  \abs{d - 1}  \sqrt{(\sigma^2_{\mathtt{clr}}+\mu_{\mathtt{clr}}^2)} = 1 \, \rightleftharpoons \, \alpha= \frac{\sqrt{2}}{\abs{d-1}  \sqrt{\sigma^2_{\mathtt{clr}}+\mu_{\mathtt{clr}}^2}}.
\end{equation}
Finally, the general expression of variability concerning $n$ observations is :
\begin{equation}
\mathtt{sCV} = \frac{\sqrt{2}}{\abs{d-1}  \sqrt{\sigma^2_{\mathtt{clr}}+\mu_{\mathtt{clr}}^2}} \sqrt{\mathbb{E}\left[(\mathtt{I}(t) - \mathtt{I}_{\mathtt{clr}}(t))^2\right]}. \end{equation}
Which is equivalent to the form slightly clearer in showing the ratio between the variability of $\mathbb{I}$ and the ``scale'' of the clear-sky irradiance:
\begin{equation}
\forall d \in [0,1[, \quad \mathtt{sCV} = \frac{\sqrt{2}}{|d-1|}\frac{\sqrt{\frac{1}{n} \sum_{i=1}^n \left( \mathtt{I}(t_i) - \mathtt{I}_{\mathtt{clr}}(t_i) \right)^2}}{\sqrt{\sigma^2_{\mathtt{clr}}+\mu_{\mathtt{clr}}^2}}
\end{equation}
Unlike the classical $\mathtt{CV}$, which reaches its maximum when the measure deviates most from the mean, this coefficient is minimized when the measure corresponds to clear-sky conditions, as it is defined relative to a dynamic upper bound rather than the mean. For competent readers, the term 
$
\sqrt{\frac{1}{n} \sum_{i=1}^n \left( \mathtt{I}(t_i) - \mathtt{I}_{\mathtt{clr}}(t_i) \right)^2}
$
represents the $\mathtt{RMSE}$ of the clear-sky model predictor $\mathtt{CS}$, which estimates $\mathtt{I}$ under clear-sky conditions.
This is the final expression for $\mathtt{sCV}$, which provides a normalized, dimensionless measure of variability. The choice of $\alpha$ ensures consistency across different scenarios and time horizons, while $\mathtt{sCV}$ isolates stochastic fluctuations ($\mathtt{I} - \mathtt{I}_{\mathtt{clr}}$) and allows for meaningful comparisons of solar resource variability. The definition of $ \mathtt{sCV} $ considers periods of constant cloud cover (\textit{e.g.}, when $ \mathtt{I} = d \times \mathtt{I}_{\mathtt{clr}} $) as highly variable, even though there are no actual changes. The hypothesis is formulated as
$\mathtt{sCV} \to 0 \implies \left| \mathtt{I} - \mathtt{I}_{\mathtt{clr}} \right| \ll \left| \mathtt{I} - d \times \mathtt{I}_{\mathtt{clr}} \right|$. In reality, cases outside this hypothesis are rare (perhaps a few periods per year where cloud cover remains high and constant for more than a day). Since these cases are negligible, previous calculations remain valid over long periods (at least one year), which simplifies the formal analysis.  
On the other hand, during long months (between spring to autumn), when cloud cover tends to be stable, $ \mathtt{I} $ remains close to $ \mathtt{I}_{\mathtt{clr}} $. A discussion about this assumption is available in Conclusion \cyr{section}. For very short time steps (\textit{e.g.}, 1 second or below), $\mathtt{sCV}$ becomes unreliable due to instrumental and physical constraints. Pyranometers and photodiodes integrate irradiance over a finite response time, acting as low-pass filters with an impulse response approximated by a sinc function. This leads to a smoothing effect that prevents the full observation of rapid transitions between clear and diffuse conditions. Additionally, at high frequencies, overirradiance effects caused by cloud-edge scattering become prominent, artificially increasing variability. From a signal processing perspective, if the sampling frequency approaches the sensor's cutoff frequency, aliasing effects may further distort the measured fluctuations. To ensure that metric correctly reflects real atmospheric variability rather than sensor-induced artifacts, a minimum time step of a few minutes is recommended, where these transient effects are naturally mitigated.

\cyr{
\subsection{Determination of $d$}
The diffuse fraction of $\mathtt{I}$ varies depending on meteorological, geographical, and atmospheric conditions. The parameter $d$, representing the diffuse fraction, is typically defined \cite{gueymard2005} by  
\begin{equation}
    d = \frac{\mathtt{I_{diff}}}{\mathtt{I}}=\frac{\mathtt{I-I_{beam}}}{\mathtt{I}},
\end{equation}  
where $\mathtt{I_{diff}}$ and $\mathtt{I_{beam}}$ are respectively the Diffuse and Beam Horizontal Irradiances \cite{duffie2013solar}. Typical values for $d$ vary with weather conditions. For convenience and consistency in variability analysis, a fixed value $d = 0.2$ is adopted \cite{SUN2021110087}, corresponding to clear-sky conditions where the diffuse fraction is minimal \cite{ruizarias2023}. While $\mathtt{sCV}$ is theoretically dependent on $d$, denoted as $\mathtt{sCV}_d$, the notation and computation by assuming $\mathtt{sCV} = \mathtt{sCV}_{0.2}$ is adopted. For instance, in certain countries, such as the United Arab Emirates, where high humidity combines with desert aerosols, the diffuse fraction can significantly exceed 0.2. Between $d = 0.2$ and $d = 0.3$, the relative difference in $\frac{1}{|d-1|}$ is approximately 14.3\%, while between $d = 0.1$ and $d = 0.2$, the difference is around 12.5\%. These variations illustrate that while $d$ influences the absolute magnitude of the metric, it does not affect the variability trends. Although this dependence on $d$ can be seen as a limitation, sensitivity analyses indicate that variations in $d$ induce changes in $\mathtt{sCV}$ of less than 15\,\%. Consequently, adopting a fixed $d=0.2$ offers a consistent and practical basis for comparing variability across different sites. It is also worth mentioning that $d$ can vary significantly during sunrise and sunset periods, due to rapid changes in the solar path length and atmospheric conditions. However, these transitions typically last less than one hour per day in total and thus have a negligible impact on long-term variability assessments.}

\subsection{Theoretical Foundations and Robustness}
The stochastic coefficient of variation ($\mathtt{sCV}$) is designed to be a part of global methods, robust and adaptable across various temporal and statistical contexts. Its definition relative to $\mathtt{I}_{\mathtt{clr}}(t)$, a dynamic upper bound, ensures independence from the observation time horizon. \cyr{It doesn’t need fixed assumptions about stationarity or homogeneity in the data, and unlike traditional measures (\textit{e.g.}, $\mathtt{CV}$) that depend on the mean $\mu$, this framework replaces the mean with $\mathtt{I}_{\mathtt{clr}}(t)$, a physically defined trend. However, it does rely on a mild assumption that the mean and variance of the sub-sampled clear-sky data (used in Eq.~6) are representative of the full period—an assumption generally valid for fine-resolution data due to the periodic and symmetric nature of clear-sky irradiance around solar noon}. This adjustment transforms the variability metric into a ``dispersion relative to a dynamic reference'', making it more suitable for bounded and periodic signals such as solar irradiance or $\mathtt{PV}$ production.
Mathematically, $\mathtt{I}(t)$ can be modeled as a stochastic process with $\mathtt{I}_{\mathtt{clr}}(t)$ acting as a deterministic upper envelope. Moments are defined relative to this reference as $m_k(\mathtt{I}) = \mathbb{E}\left[(\mathtt{I}(t) - \mathtt{I}_{\mathtt{clr}}(t))^k\right],$
where $k = 2$ corresponds to the variance relative to $\mathtt{I}_{\mathtt{clr}}(t)$. This probabilistic approach avoids the constraints of strict stationarity, as the variability is quantified with respect to a time-varying reference \cite{GORDON198935}. Consequently, the measure remains robust under heteroscedastic conditions, where the variance changes dynamically over time.
Other theoretical concepts, such as linearity, are available in \ref{properties} and an extension to the volatility concept is available in \ref{volatility}.

\subsection{Link with Predictability and Forecastability}
Variability quantifies the magnitude of stochastic fluctuations around a deterministic trend, while forecastability $\mathtt{F}$ measures how well these fluctuations can be anticipated, incorporating both variability and temporal dependencies. Predictability, linked to entropy and autocorrelation, provides the theoretical foundation for forecastability but remains distinct \cite{MAASOUMI2002291,e25111542}.
For cyclostationary signals \cite{https://doi.org/10.1002/env.2700}, the temporal structure of residuals $(\epsilon = \mathtt{I} - \mathtt{I}_{\mathtt{clr}})$ is key to defining forecastability. This structure is captured by the maximum autocorrelation $\rho_{\max}$ of the residuals over lags $i$, where $i \in [n , T / 2 ]$, capturing temporal dependencies over this interval. Here, $T$ is the number of time steps in one period of the signal. (\textit{e.g.} $T=24$ for hourly data over a day in the solar case), and $n$ corresponds to the lag horizon (\textit{e.g.} $n=6$ for 6-h horizon in hourly data). The total horizon in physical time is given by $n \Delta t$ where $\Delta t$ is the time step between successive observations. The autocorrelation at lag $i$ is defined as:
\begin{equation}
\rho\big(\mathtt{\epsilon}(t), \mathtt{\epsilon}(t - i \Delta t)\big) =\rho (i) = \frac{\text{Cov}\big(\mathtt{\epsilon}(t), \mathtt{\epsilon}(t - i \Delta t)\big)}{\sigma(\mathtt{\epsilon})^2},
\end{equation}
where $\text{Cov}(\cdot)$ denotes the covariance, and $\sigma(\mathtt{\epsilon})$ is the standard deviation of $\mathtt{\epsilon}$. The proposed metric for forecastability is then, considering $\rho_{max} = \max_{i \in [n , T / 2 ]} \rho(i)$:
\begin{equation}
\label{eq:rhomax}
\mathtt{F} = (1 - \mathtt{sCV}) +\rho_{max} \mathtt{sCV}   = 1 + \mathtt{sCV} \big( \rho_{max} -1 \big).
\end{equation}
\cyr{Note that although $\rho(i)$ can be negative for some lags $i$, $\rho_{\max}$ is defined as the maximum absolute value of the autocorrelation function over the considered lags, so it is always $\geq 0$}. This formulation ensures that $\mathtt{F}$ is bounded within $[1-\mathtt{sCV}, 1]$. When $\rho_{\max} = 0$, $\mathtt{F} = 1- \mathtt{sCV}$, reflecting the absence of temporal dependencies. When $\rho_{\max} = 1$, $\mathtt{F} = 1$, indicating perfect forecastability regardless of variability. For intermediate values of $\rho_{\max}$, $\mathtt{F}$ interpolates smoothly between variability and the upper bound of forecastability.
This metric aligns naturally with the cyclostationary nature of solar irradiance. Temporal dependencies captured by $\rho_{\max}$ enhance predictability, allowing forecastability to exceed variability in structured signals \cite{karimi2023quantifying}. For example, in periodic signals with strong autocorrelations, $\mathtt{F}$ reflects both the magnitude of fluctuations ($\mathtt{sCV}$) and their temporal coherence ($\rho_{\max}$). Conversely, in unstructured signals where $\rho_{\max} \approx 0$, forecastability reduces to the complementary of variability ($1-\mathtt{sCV}$), as no temporal patterns aid prediction. If $\mathtt{sCV}$ is independent of the time horizon and the inclination of solar irradiance, $\mathtt{F}$ depends on it according to the use in its definition of the autocorrelation $\rho(i)$.
The forecastability metric $\mathtt{F}$ is conceptually similar to the coefficient of determination $\mathtt{R^2}$ in regression analysis. While $\mathtt{R^2}$ measures the proportion of variance explained by an external model, $\mathtt{F}$ quantifies the proportion of variability in a time series that can be anticipated based on its temporal dependencies.
Thus, $\mathtt{F}$ integrates stochastic variability and temporal dependencies, much like $\mathtt{R^2}$ combines residuals and explained variance, providing a unified framework for predictability.

\subsection{Methods of Prediction}
The prediction of solar irradiance relies on a variety of methods, ranging from simple empirical techniques to advanced machine learning models. Among the most basic approaches, persistence ($\mathtt{P}$ \citep{kumler2019smartpersistence}) assumes that the future irradiance remains the same as the most recent observation. This method serves as a standard benchmark due to its simplicity and effectiveness for very short-term forecasts.
A refinement of this approach is the smart persistence model ($\mathtt{P_S}$ \citep{Perez1990}), which incorporates clear-sky calculation with, in this case, an additive rather than multiplicative mode using $\mathtt{k_c}$ to avoid problems during sunshine, sunset and night. This method adjusts for diurnal variations and offers improved accuracy under clear and partially cloudy conditions.
Periodic autoregressive ($\mathtt{PAR}$ \citep{VOYANT2018121}) models extend the capabilities of standard autoregressive ($\mathtt{AR}$) models by including periodic components, making them particularly effective for capturing seasonal and daily patterns in irradiance.
\cyr{Extreme Learning Machines ($\mathtt{ELM}$ \citep{8074179}) offer a distinct approach by using a single hidden layer feedforward neural network. Unlike deep learning models, $\mathtt{ELM}$ is not ``deep'' but instead focus on computational efficiency and fast training. It achieve this by fixing the weights between the input and hidden layers randomly and only optimizing the output weights analytically (closed-form). $\mathtt{ELM}$ is also capable of modeling nonlinear relationships due to the activation functions in its hidden layer, making it particularly suitable for capturing the stochastic and complex dynamics of solar irradiance.}
These methods provide a broad spectrum of predictive capabilities, from simple benchmarks to sophisticated models capable of capturing both stochastic fluctuations and nonlinear dynamics in solar irradiance data.
\section{Results}
\label{sec:results}
This section presents and analyzes results obtained from synthetic time series with controlled characteristics \cite{POLO20111164}, as well as experimental measurements collected from 68 stations in the $\mathtt{SIAR}$ network \cite{DESPOTOVIC2024123215} with ad-hoc Quality Check \cite{app12178529}. These analyses aim to validate the robustness and applicability of the proposed variability metrics.

\subsection{General Framework in Synthetic Series Context}
\label{sec:synthetic}
To ensure the robustness of variability metrics in different scenarios, a Monte Carlo simulation framework capable of generating diverse ensembles of synthetic time series \cyr{is proposed} \cite{Voyant2021}. These synthetic series rely on cyclostationary processes, which are ideal for modeling solar irradiance dynamics thanks to their periodic statistical properties \cite{Shevgunov_2019,syntheticintro}.
The generation process starts with a deterministic periodic component that represents the clear-sky solar cycle, characterized by predictable diurnal patterns in solar irradiance. This baseline is modeled as a sine wave adjusted to ensure non-negativity, capturing the smooth diurnal variation of solar irradiance. Gaussian noise is superimposed onto the periodic signal to simulate stochastic variability introduced by atmospheric conditions. 
To replicate smooth transitions typical of experimental irradiance, temporal correlations are introduced by applying a moving average filter to the noisy signal. The filter length is a key parameter controlling the smoothness of the signal; shorter filter lengths retain high-frequency variability, while longer filters produce smoother profiles.
Monte Carlo simulations were conducted by randomly sampling key parameters from predefined ranges to ensure a diverse set of synthetic time series \cite{Voyant2021}. The period $T$ corresponds to the repeating pattern of the solar cycle, typically representing a 24-hour day. For these simulations, $T = 240$ points per period corresponds to a temporal resolution of six minutes per time step. \cyr{Here, the product $n \Delta t$ defines the total time horizon (or total duration), where $n$ is the total number of futur data points (lag horizon) and $\Delta t$ is the time step. Although the use of $n$ occur in different parts of the manuscript, they represent a lag or shift in the data, whether backward (past lags) or forward (forecast horizon).}
A total of 100 synthetic time series were generated, each with unique parameter configurations. Due to the large number of points, displaying all results hinders readability. Therefore, either averages or a representative sample of 50 time series \cyr{is presented}, which strikes a good balance between detail and clarity. The randomization of parameters, such as noise amplitude, filter length, and temporal resolution, ensures that the ensemble captures the diversity and variability inherent in experimental solar irradiance dynamics. 
This tunable simulation framework enables systematic exploration of metric sensitivity and robustness under controlled yet realistic conditions. These synthetic series not only facilitate controlled sensitivity analyses but also provide a solid foundation for comparing variability metrics in experimental applications.

\subsubsection{Evolution of the Parameters According to the Horizon with Synthetic Series}
\begin{figure}[ht]
    \hspace{-0.07\textwidth} 
    \includegraphics[width=1.15\textwidth]{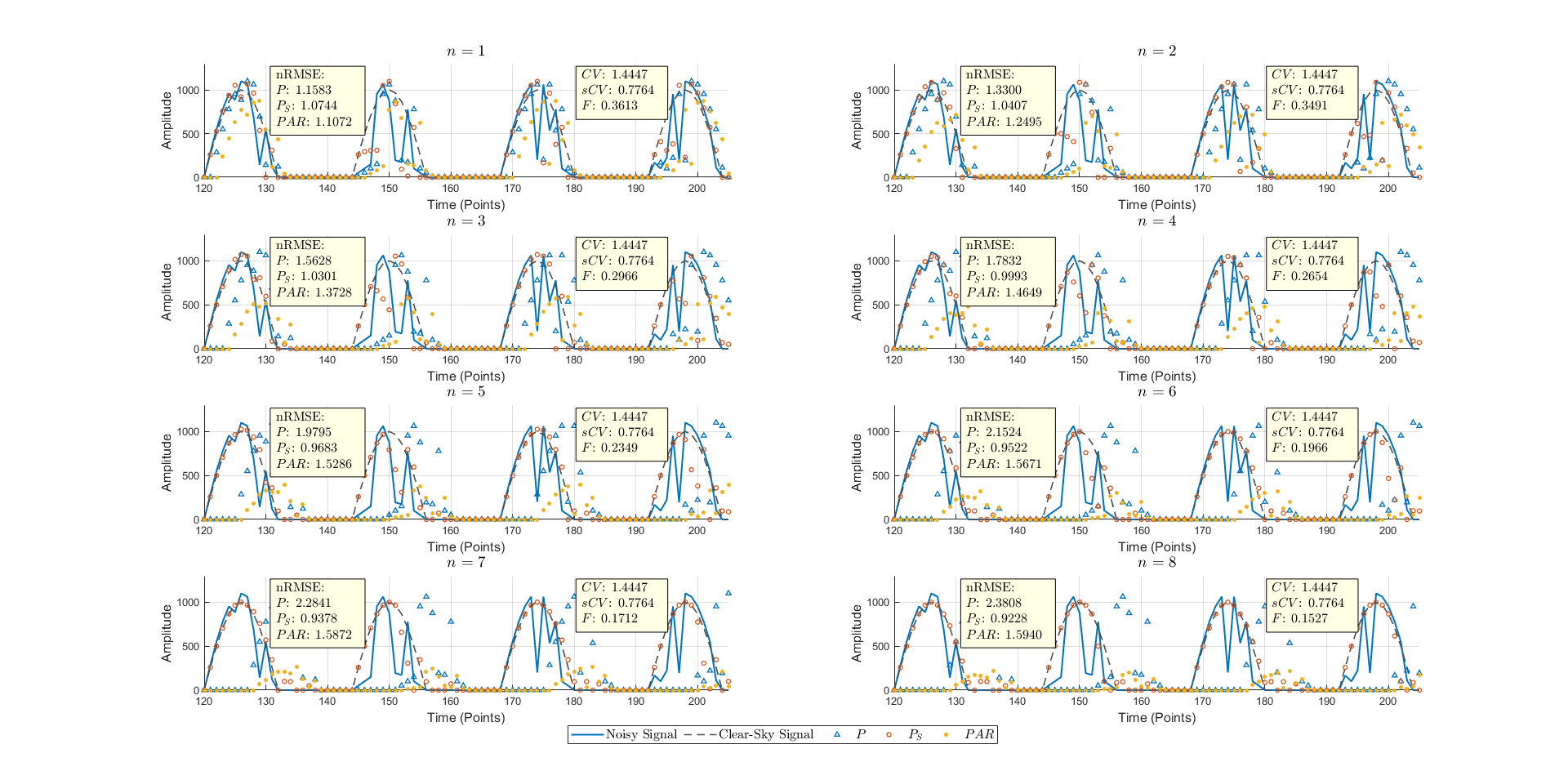}
    \caption{
       Prediction models ($\mathtt{P}$, $\mathtt{P}_\mathtt{S}$, $\mathtt{PAR}$) compared for different time horizon ($n\Delta t$, where $n$ is the lag horizon). 24 elements are considered per period ($T=24$) with $\Delta t=1$. Each subplot shows the noisy signal, clear-sky signal, model predictions, and metrics ($\mathtt{nRMSE}$, $\mathtt{CV}$, $\mathtt{sCV}$, $\mathtt{F}$).
           }
    \label{fig:profiles_subplots}
\end{figure}

\cyr{Figure~\ref{fig:profiles_subplots} illustrates the variability ($\mathtt{CV}$, $\mathtt{sCV}$) and forecastability ($\mathtt{F}$) of the signals across different lag horizons ($n \in [1,8]$) and prediction models ($\mathtt{P}$, $\mathtt{P}_\mathtt{S}$, $\mathtt{PAR}$).}  
The coefficient of variation ($\mathtt{CV}$) remains constant at $1.4447$, indicating a fixed relative variability of the noisy signals. In contrast, the stochastic coefficient of variation ($\mathtt{sCV}$), which captures stochastic fluctuations around the clear-sky signal, stabilizes at $0.7764$, reflecting consistent deviations irrespective of the time horizon.  
Forecastability ($\mathtt{F}$), \cyr{however, decreases with increasing $n$, dropping from $0.3613$ to $0.1527$}. This trend highlights the progressive loss of temporal structure as the horizon extends, consistent with the reduction of $\rho_{\max}$ (maximum autocorrelation) at longer horizons.  
\cyr{Additionally, the root mean squared error ($\mathtt{RMSE}$), when normalized by the mean of the measured signal (normalized $\mathtt{RMSE}$, $\mathtt{nRMSE}$), 
demonstrates a clear relationship with $\mathtt{F}$ but not necessarily with $\mathtt{sCV}$. 
This suggests that the prediction error is more closely linked to the forecastability measure ($\mathtt{F}$), rather than directly to the variability metric ($\mathtt{sCV}$). 
It is important to note that while $\mathtt{nRMSE}$ provides a relative error measure, it does not necessarily remain below 1. 
Indeed, $\mathtt{nRMSE}$ can exceed 1 when the RMSE is larger than the mean of the measured signal.
These results underscore the importance of using $\mathtt{F}$ and $\mathtt{sCV}$ as complementary metrics for assessing predictability in variable data}.

\subsubsection{Robustness with Synthetic Series}
\begin{figure}[h]
    \centering
    \includegraphics[width=0.9\textwidth]{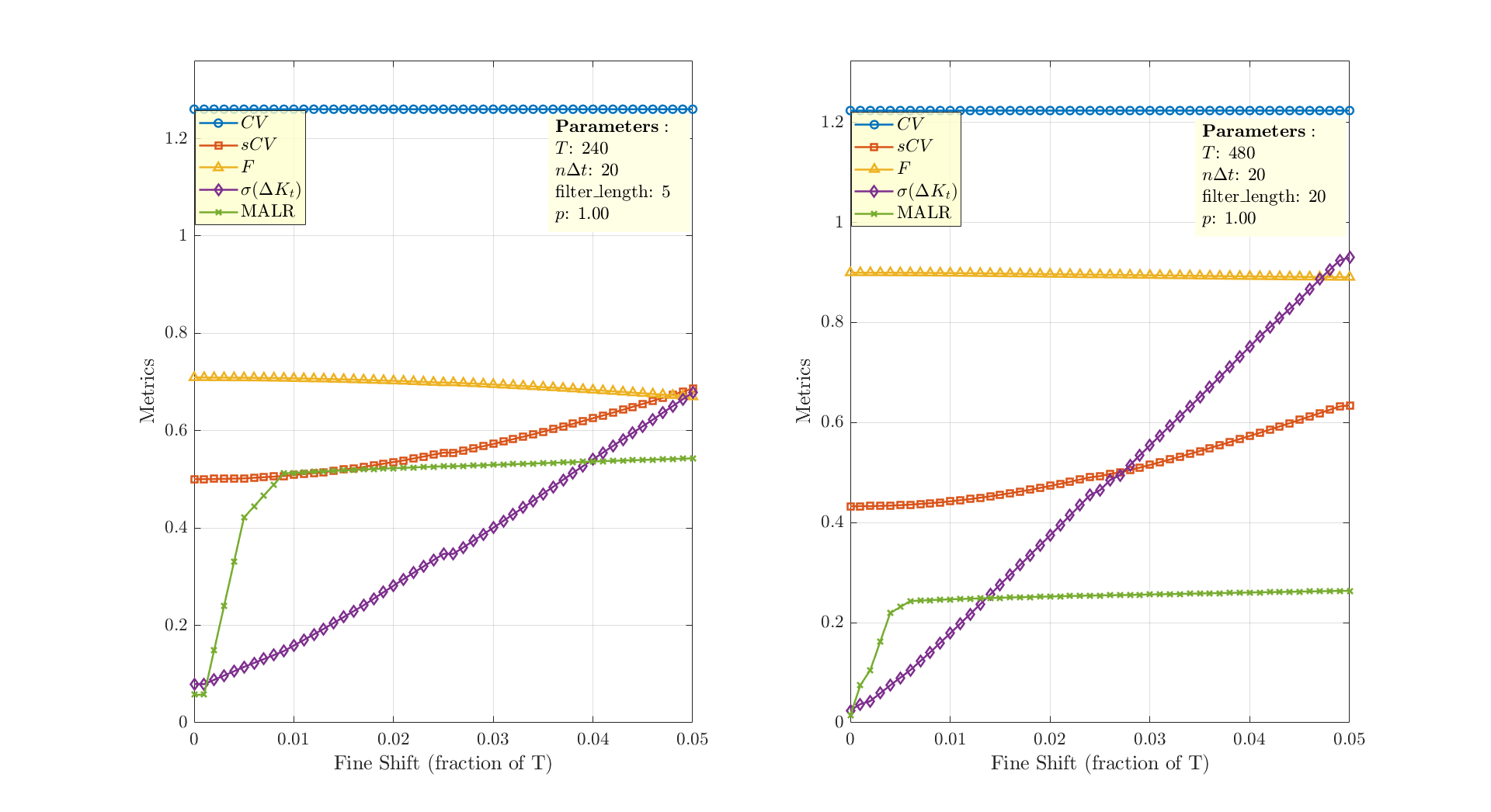}
    \caption{Impact of fine phase shifts on variability and forecastability metrics}
    \label{fig:phase_off}
\end{figure}

Figure~\ref{fig:phase_off} illustrates the impact of phase shifts (expressed as a fraction of the clear-sky periodicity $\mathtt{T}$) on variability and forecastability metrics. If $T=24$ so $5\%$ of $T$ $(=0.05T)$ represent slight more than 1h, in the real case concerning $\mathtt{I}$, the shift is less than 20-30 minutes. The left panel corresponds to $240$ points per period with a moving average filter length of $5$, while the right panel uses $480$ points per period with a filter length of $20$.
The results show that $\mathtt{CV}$ remains constant between $1.2-1.3$ in the two cases, demonstrating its robustness to phase shifts. In contrast, $\mathtt{sCV}$ increases moderately by $38\%$ (\textit{e.g.}, from $0.5$ to $0.7$ for $T=240$ points), while $\mathtt{F}$ varies minimally (around $5\%$, \textit{e.g.}, from $0.70$ to $0.65$). These findings confirm the stability of these metrics against misalignments.
In comparison, $\sigma(\Delta \mathtt{k_c})$ and $\mathtt{MALR}$ are highly sensitive to phase shifts. For $T=240$ points, $\sigma(\Delta \mathtt{k_c})$ increases by $273\%$ (from $0.07$ to $0.7$) and $\mathtt{MALR}$ by $88\%$ (from $0.05$ to $0.5$). Similar trends are observed for $T=480$ points, with increases of $286\%$ and $88\%$, respectively. 
These results highlight that $\mathtt{CV}$ and $\mathtt{F}$ are robust metrics, suitable for analyzing signals with minor misalignments. However, metrics like $\sigma(\Delta \mathtt{k_c})$ and $\mathtt{MALR}$ require careful calibration due to their significant sensitivity to phase shifts. The moderate increase in $\mathtt{sCV}$ ($38\%$) makes it a reliable indicator of broader variability trends.

\subsubsection{Benchmark with Classical Predictors}
\begin{figure}[H]  
    \hspace{-0.1\textwidth} 
    \includegraphics[width=1.15\textwidth]{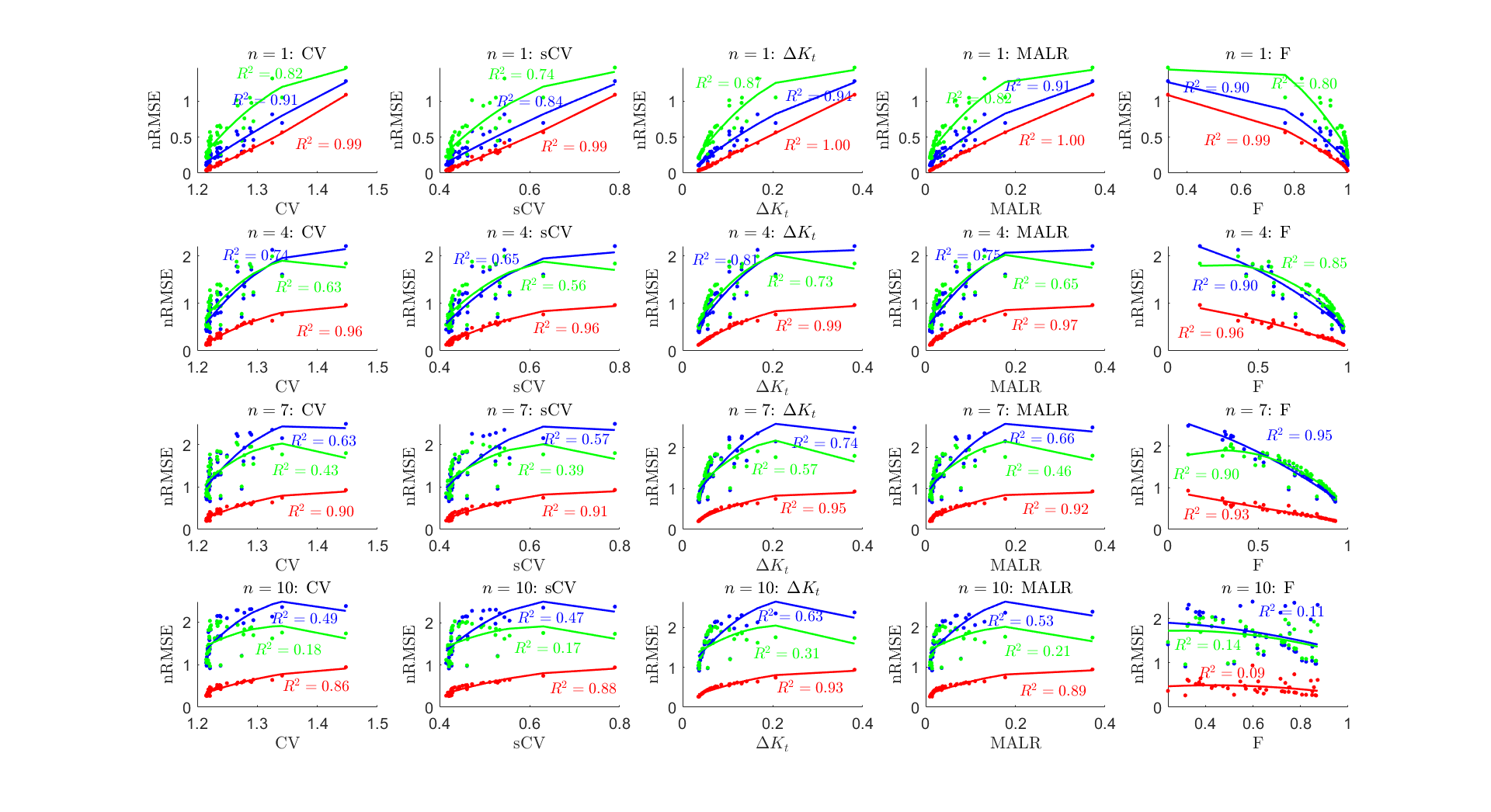} 
    \caption{Scatter plots of $\mathtt{nRMSE}$ as a function of variability and forecastability metrics ($\mathtt{CV}$, $\mathtt{sCV}$, $\mathtt{\Delta k_c}$, $\mathtt{MALR}$, and $\mathtt{F}$) for different prediction lag horizons $ = \{1, 4, 7, 10\}$ and synthetic cyclostationary processes. Each subplot displays scatter points for the three prediction models ($\mathtt{P}$, $\mathtt{P}_\mathtt{S}$, $\mathtt{PAR}$) in blue, red, and green, respectively, along with a second-degree polynomial trend line for each model. The corresponding $\mathtt{R}^2$ values quantify the quality of the fit.}
    \label{fig:MonteCarloResults} 
\end{figure}

The relationship between normalized Root Mean Squared Error ($\mathtt{nRMSE}$) and metrics of variability and forecastability was analyzed for prediction lag horizons $n = \{1, 4, 7, 10\}$, as shown in Figure~\ref{fig:MonteCarloResults}. Coefficient of Variation ($\mathtt{CV}$), stochastic Coefficient of Variation ($\mathtt{sCV}$), intermittency ($\mathtt{\Delta k_c}$), Mean Absolute Log Return ($\mathtt{MALR}$), and Forecastability ($\mathtt{F}$) are presented. Among these metrics, forecastability ($\mathtt{F}$) exhibited the strongest correlation with $\mathtt{nRMSE}$, as evidenced by consistently high $\mathrm{R}^2$ values and monotonic trends across all horizons. To statistically validate this relationship, a Spearman rank correlation test was conducted. The results confirmed significant positive correlations at all horizons: $n = 1$ ($\rho = 0.89$, $p = 0.001$), $n = 4$ ($\rho = 0.83$, $p = 0.002$), $n = 7$ ($\rho = 0.78$, $p = 0.007$), and $n = 10$ ($\rho = 0.65$, $p = 0.025$). These findings highlight the robustness of $\mathtt{F}$ as a predictor of forecast accuracy. 
Monotonicity, rather than linearity, is critical in this analysis. A monotonic relationship ensures that as $\mathtt{F}$ decreases, prediction error ($\mathtt{nRMSE}$) increases, even if the relationship is not strictly linear. This property is essential for practical applications in energy forecasting and is supported by results, making $\mathtt{F}$ a valuable tool for benchmarking sites.
In contrast, variability metrics such as $\mathtt{CV}$ and $\mathtt{sCV}$, while capturing specific aspects of data variability, showed weaker correlations and lower $\mathtt{R^2}$ values, especially for longer horizons. This is likely because these metrics focus on inherent variability within the time series itself rather than directly addressing the predictability of future values. Moreover, they may be more sensitive to short-term fluctuations, which are less relevant for forecasting accuracy at extended horizons.
These results establish $\mathtt{F}$ as the most reliable metric for assessing predictability. Its robustness across prediction models ($\mathtt{P}$, $\mathtt{P_S}$, $\mathtt{PAR}$) and independence from the specific forecasting method make it particularly suitable for practical applications.

\subsection{\texttt{SIAR} Data}
\label{sec:siar_data}
\begin{figure}[ht]
    \centering
    \includegraphics[width=0.7\linewidth]{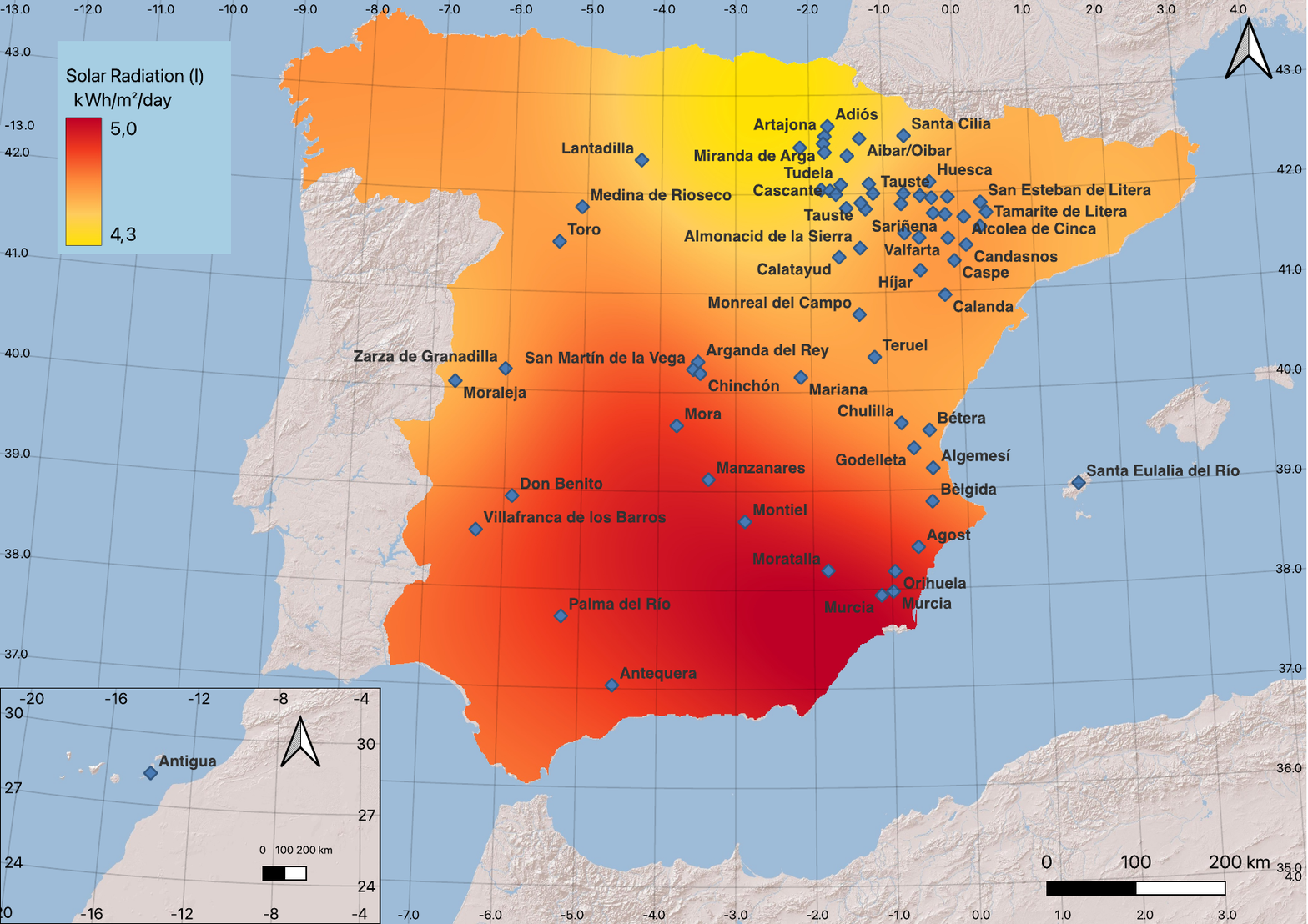}
    \caption{\cyr{Geographical distribution of $\mathtt{SIAR}$ stations and kriging surface of daytime-averaged $\mathtt{I}$ (2017-2020)}} 
    \label{fig:siar_map}
\end{figure}

While synthetic time series offer valuable insights into the behavior and robustness of the variability ($\mathtt{sCV}$) and forecastability ($\mathtt{F}$) metrics, it is essential to validate these results under real atmospheric conditions. To this end, the next section focuses on an experimental evaluation using data from the $\mathtt{SIAR}$ network, which includes 68 geographically distributed stations across Spain. This validation step involves applying the full framework on real irradiance time series: variability computation, forecastability analysis, and model performance evaluation. The goal is to test whether the relationships observed in synthetic simulations (particularly the link between $\mathtt{F}$ and prediction error metrics) hold across diverse climatic conditions. 
Forecast accuracy is assessed using $\mathtt{RMSE}$ across 10 forecasting models. These models, previously detailed in the benchmark study \cite{voyant_benchmarks_2022}, include: autoregressive model ($\mathtt{AR}$), smart persistence ($\mathtt{SP}$), standard persistence ($\mathtt{P}$), clear-sky model ($\mathtt{CS}$), climatology-persistence optimal combination ($\mathtt{CLIPER}$), exponential smoothing ($\mathtt{ES}$), particular autoregressive model of order two ($\mathtt{ARTU}$), and combination of models ($\mathtt{COMB}$). In addition, the two forecasting models introduced in Section~\ref{sec:methodology} extreme learning machine ($\mathtt{ELM}$) and periodic autoregressive ($\mathtt{PAR}$), are included.
\cyr{The irrigation agroclimatic information system (\textit{i.e.}, Sistema de Información Agroclimática para el Regadío $\mathtt{SIAR}$), managed by the Spanish state meteorological agency (\textit{i.e.}, Agencia Estatal de Meteorología $\mathtt{AEMET}$), collects $\mathtt{I}$ data using calibrated pyranometers at stations distributed throughout Spain}. These sites span diverse geographic and climatic conditions, making the dataset well-suited for analyzing solar irradiance variability. The network follows established calibration and quality control protocols to ensure data reliability. Figure \ref{fig:siar_map} displays the locations of the 68 stations included in this study, illustrating their coverage of distinct solar irradiation regions.
In addition to station locations, Figure \ref{fig:siar_map} incorporates a geostatistical interpolation method (ordinary kriging) to produce a spatially continuous map of daytime-averaged $\mathtt{I}$. Ordinary kriging estimates the irradiance $\mathtt{I}^{*}(\mathbf{x}_0)$ at an unsampled location $\mathbf{x}_0$ as a weighted linear combination of observations from surrounding stations $\{\mathbf{x}_i\}$, with $\mathtt{I}^{*}(\mathbf{x}_0) \;=\; \sum_{i=1}^{n} \lambda_i \, \mathtt{I}(\mathbf{x}_i),$ considering that $\sum_{i=1}^{n} \lambda_i \;=\; 1$ and $\mathtt{I}(\mathbf{x}_i)$ is the measured $\mathtt{I}$ at station $\mathbf{x}_i$, $n$ is the number of nearby stations, and $\lambda_i$ are weights derived from semivariogram \cite{pereira2022smart}.  \cyr{Although the study includes 68 stations, the interpolation shown in the figure was done using the 66 located on the mainland. The two island-based stations, in the Balearic Islands \cyr{(Santa Eulalia del Rio Station)} and Canary Islands (Antigua Station), were excluded due to their geographic separation from mainland Spain \cyr{(see Figure \ref{fig:siar_map})}.}

\subsubsection{General Framework of \texttt{SIAR} Data}
\label{sec:siar_validation}
The dataset includes both the measured global horizontal irradiance, denoted as $\mathtt{I}(t)$, and the corresponding clear-sky irradiance $\mathtt{I_{clr}}(t)$. 
The period $T$ corresponds to the repeating pattern of the solar cycle, typically representing a 24-hour day. For these simulations, $T = 48$ points per period corresponds to a temporal resolution of 30 minutes per time step. Ten different forecasting models are evaluated concerning horizons ranging from $n=1$ (30 minutes ahead) to $n=24$ (12 hours ahead). 

\subsubsection{Autocorrelation and Forecastability}
\begin{figure}[H]
    \centering
    \includegraphics[width=0.8\linewidth]{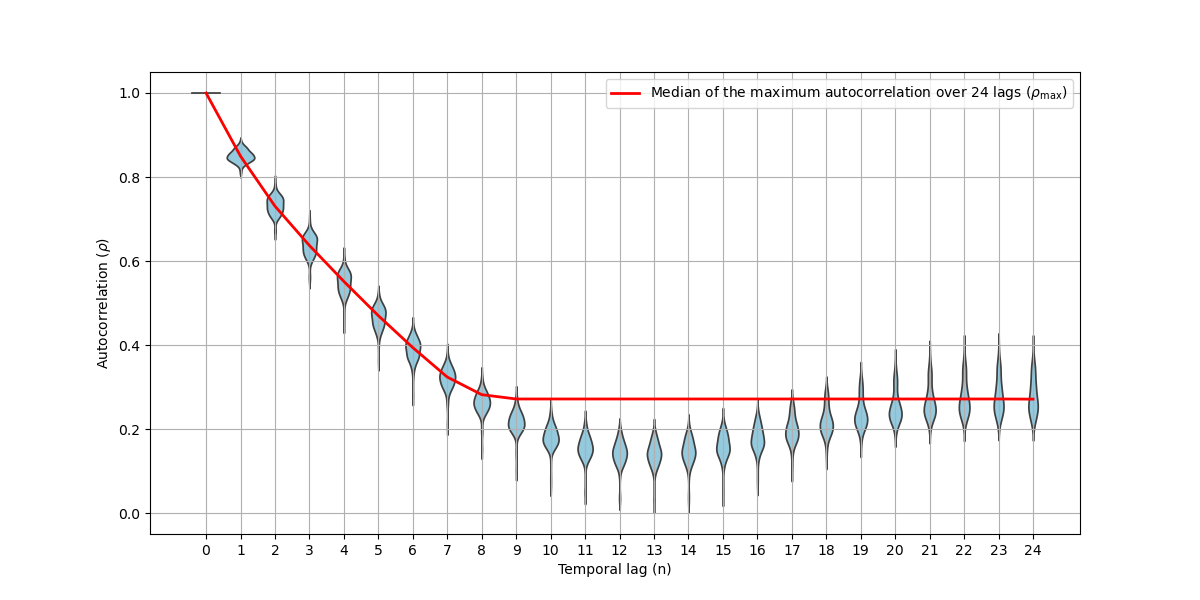}
    \caption{Violin boxplot of auto-correlationn factor ($\mathtt{ACF}$) $\rho$ across 68 stations for forecast horizons $n = 1$ to $24$ (30min to 12h). The red line shows the median $\rho_{\max}$ across all cases. A distinct inflection occurs at $n = 8$, indicating a structural threshold in forecastability.}
    \label{fig:rho}
\end{figure}

This section analyzes the autocorrelation of solar irradiance data across all stations, examining various forecasting horizons. To connect autocorrelation ($\mathtt{ACF}$) with forecastability, the maximum correlation coefficient (denoted as $\rho_{\max}$) is computed over a time window of $T/2$, corresponding to 24 forecasting horizons (see Eq. \ref{eq:rhomax}). 
Violin boxplots (Figure~\ref{fig:rho}) visualize the distribution of the $\mathtt{ACF}$ for lags ranging from $n = 1$ to $n = 24$, highlighting the variation of $\rho$ at each horizon. The median $\rho_{\max}$ across all stations is shown as a red line.
The analysis reveals a noticeable change of the $\rho_{\max}$ red line in slope at $n = 8$, suggesting a structural shift in $\mathtt{F}$. Beyond $n = 9$, $\rho_{\max}$ remains constant, as expected due to the nature of the $\mathtt{F}$ metric, which reflects the horizon beyond which $\mathtt{F}$ remains constant. Thus, $\mathtt{F}$ validation focuses on horizons $n = 1$ to $8$. The observed stabilization of autocorrelation beyond $n=8$ suggests the existence of an effective correlation horizon, beyond which the marginal gain in $\mathtt{F}$ becomes negligible. Unlike a strict Lyapunov horizon \footnote{Defined as the inverse of the largest Lyapunov exponent, it can detect the presence of deterministic chaos in complex systems (also in radiative processes) and quantify the predictability of potential future outcomes \cite{e20080570}} where predictability vanishes, here $\rho_{max}$ converges to a non-zero asymptote, indicating residual structure that remains statistically exploitable.

\subsubsection{Relationship Between Forecastability and Error Metrics}
\begin{figure}[H]
    \centering
    \includegraphics[width=\linewidth]{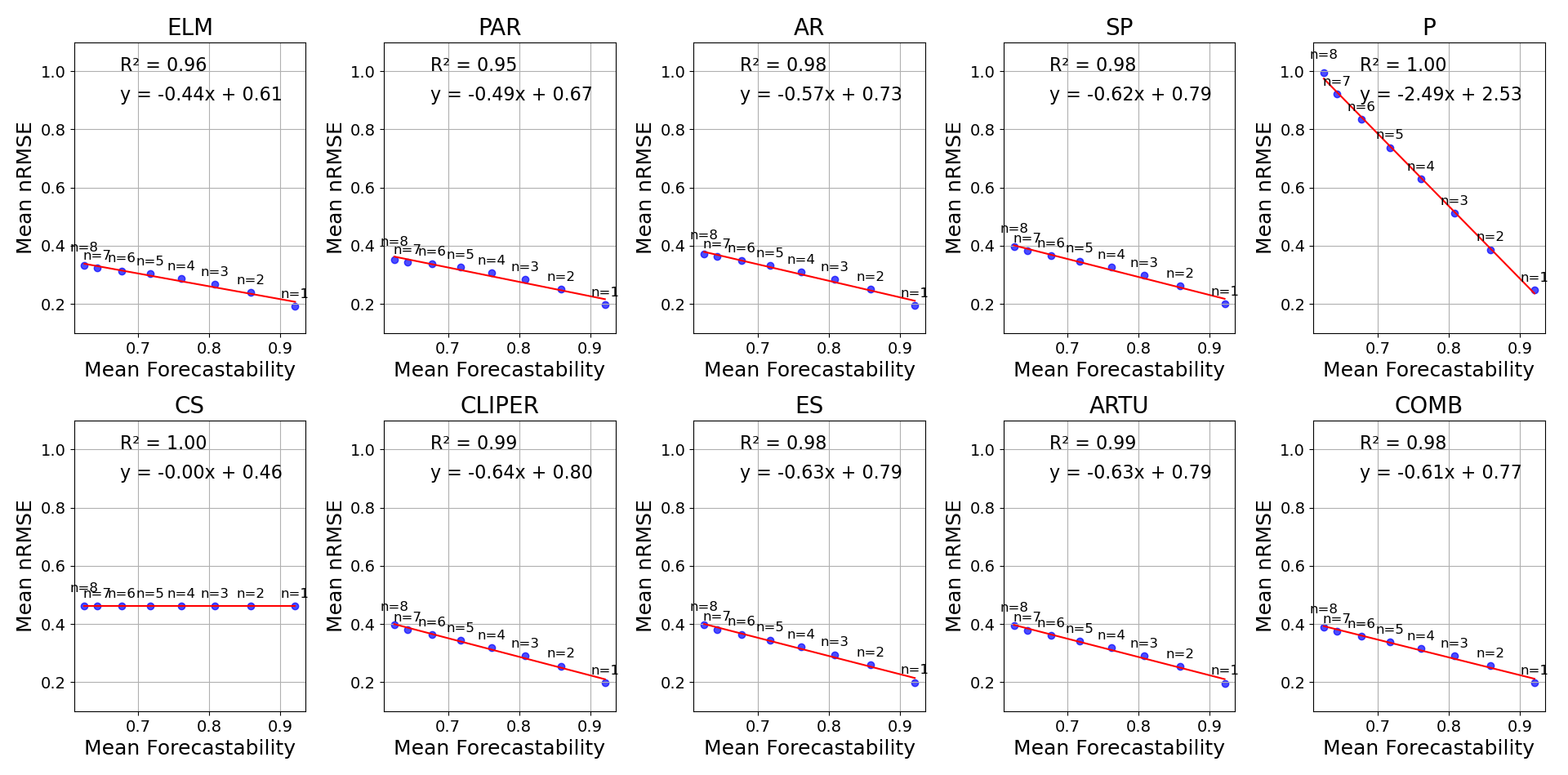}
    \caption{Scatter plots of mean forecast error ($\mathtt{nRMSE}$) versus forecastability $\mathtt{F}$ across 68 stations for 10 forecasting models over horizons $n = 1$ to $8$ indicated over data points. From $n=8$ to $n=24$, the points are merged (see the $\rho$ singularity in figure \ref{fig:rho}). A linear regression and related $\mathtt{R}^2$ values quantify the relationship.}
    \label{fig:Fscatters}
\end{figure}

To further assess the relationship between $\mathtt{F}$ and model performance, 10 forecasting models have been analyzed. For each model, a scatter plot (Figure~\ref{fig:Fscatters}) shows the mean forecast error ($\mathtt{nRMSE}$) of $\mathtt{SIAR}$ stations as a function of their mean $\mathtt{F}$ value across horizons from $n = 1$ to $8$.
Results indicate a strong linear correlation (with $\mathtt{R}^2$ between 0.95 and 1.00) between $\mathtt{F}$ and $\mathtt{nRMSE}$. All models except the clear-sky model ($\mathtt{CS}$) exhibit a negative slope, demonstrating that $\mathtt{nRMSE}$ decreases as $\mathtt{F}$ increases. $\mathtt{CS}$ maintains nearly constant $\mathtt{nRMSE}$, due to its static nature. The persistence model ($\mathtt{P}$) shows higher $\mathtt{nRMSE}$ and a steeper slope, reflecting its limited predictive power. Across other models, $\mathtt{nRMSE}$ values vary from 0.2 to 0.4 as $\mathtt{F}$ increases from 0.6 to 0.95, confirming that $\mathtt{F}$ effectively predicts forecast performance.

\subsubsection{Regression Consistency Across Models}
\begin{figure}[H]
    \centering
    \includegraphics[width=\linewidth]{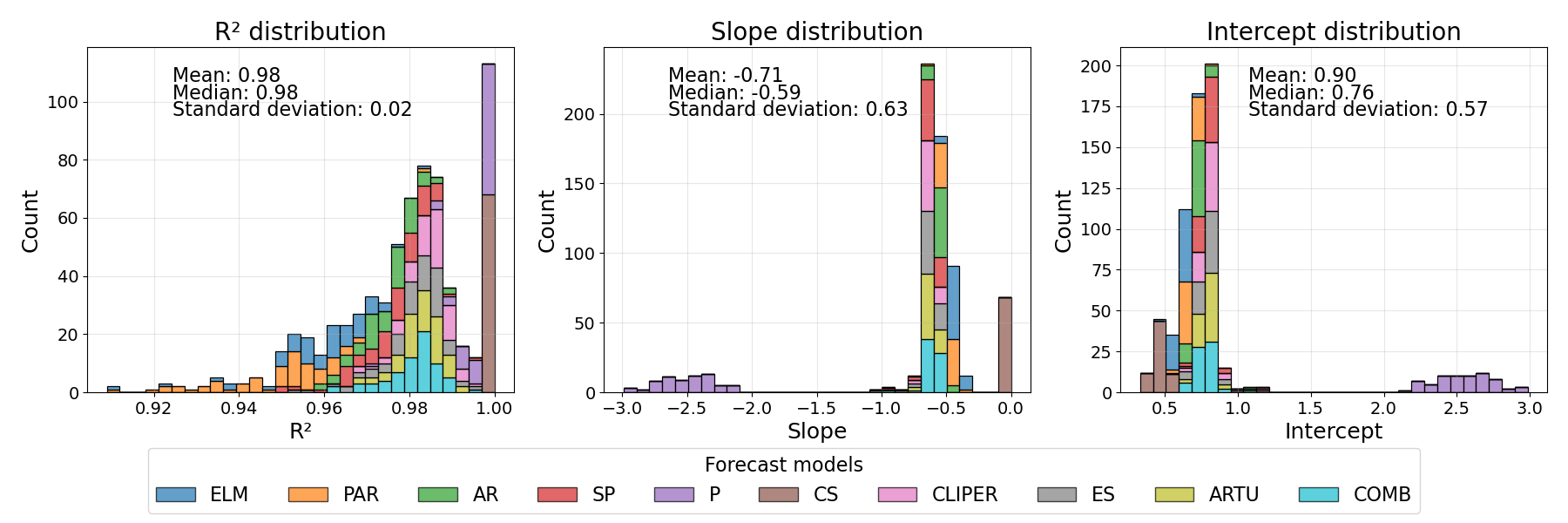}
    \caption{Distributions of regression parameters between forecastability $\mathtt{F}$ and $\mathtt{nRMSE}$ at each station: $\mathtt{R}^2$, $\mathtt{slope}$, and $\mathtt{intercept}$, aggregated by forecasting model.}
    \label{fig:distributions}
\end{figure}

To ensure that the $\mathtt{F}$–$\mathtt{nRMSE}$ relationship is consistent at individual stations, a regression analyses is performed between them for each model and each of the 68 stations. Rather than displaying all 680 regressions, Figure~\ref{fig:distributions} summarizes the distributions of slope, intercept, and $\mathtt{R}^2$ by model. 
The $\mathtt{R}^2$ values are tightly clustered between 0.90 and 1.00 (mean = 0.98), with peaks close to $\mathtt{R}^2 = 1.0$ for the $\mathtt{CS}$ and $\mathtt{P}$ models, reflecting their deterministic nature. Flexible models such as $\mathtt{ELM}$ and $\mathtt{PAR}$ show broader but still high $\mathtt{R}^2$ distributions. Slopes are predominantly negative (mean = –0.71), with a bimodal pattern: most models cluster between –0.6 to –0.8, while $\mathtt{P}$ has steeper slopes (–2.0 to –3.0). The $\mathtt{CS}$ model yields near-zero slopes due to constant error levels. Intercepts mainly range from 0.5 to 1.0, with a secondary mode around 2.5–2.9, again driven by the high baseline error of the $\mathtt{P}$ model.
\section{Conclusion and Practical Implications for Operators}
\cyr{This study introduced the $\mathtt{sCV}$ and $\mathtt{F}$ metrics, which provide a novel and rigorous framework for quantifying the variability and forecastability of solar irradiance. Unlike traditional indicators that rely on static references and cannot separate deterministic and stochastic components, $\mathtt{sCV}$ isolates stochastic fluctuations relative to the dynamic clear-sky trend, while $\mathtt{F}$ integrates this with temporal predictability through autocorrelation. These physically grounded, bounded, and horizon-independent metrics are particularly relevant for applications such as solar forecasting, grid integration, and dynamic reserve sizing, addressing key limitations of classical variability indicators and enabling more reliable decision-making in energy systems.}
In Section~\ref{sec:synthetic}, 100 synthetic cyclostationary time series were used to explore the behavior of $\mathtt{sCV}$ and $\mathtt{F}$ under controlled variability and noise conditions. These simulations confirmed that $\mathtt{sCV}$ remains stable across different horizons and signal characteristics, while $\mathtt{F}$ exhibits a strong relationship with prediction error, especially when using simple and interpretable models. The robustness of $\mathtt{F}$ was further supported by its consistent statistical dependency with $\mathtt{nRMSE}$ across models and time scales, establishing it as a reliable indicator of inherent predictability.
Section~\ref{sec:siar_data} extended this analysis to real measurement from the $\mathtt{SIAR}$ network, comprising 68 geographically distributed stations in Spain. The empirical evaluation confirmed the theoretical expectations: $\mathtt{F}$ maintained a strong inverse relationship with prediction error across 10 forecasting models and 24 time horizons. At the network level, $\mathtt{ACF}$ analyses revealed a structural horizon of predictability, while scatterplots and regression results demonstrated that $\mathtt{F}$ effectively predicts model performance. Notably, this relationship remained consistent at the station level, ruling out spatial or aggregation biases and validating the generalizability of the metric across different climatic contexts.
Finally, the linear nature of the dependency observed between $\mathtt{F}$ and forecast error in the measured data likely reflects the fact that the variability of the Spanish time series remains moderate; as shown by the synthetic series, statistical dependence becomes clearly monotonic but non-linear under extreme variability conditions.
Beyond classical forecasting applications, these metrics are particularly useful for operational decision-making in energy management. By incorporating them into forecasting models, it becomes possible to refine flexibility procurement strategies, adjusting conservatism levels based on site predictability. In particular, this methodology can support dynamic outage management, where forecast confidence informs the temporary restoration of generation capacity during planned outages. Sites with high predictability allow for greater capacity release, while uncertain locations necessitate a more cautious approach. These insights enable more efficient resource allocation, ultimately improving the economic and operational performance of energy systems.
A promising prospect would be to adapt this approach in on-line mode, in order to offer an intra-day tool based on $\mathtt{sCV}$ and $\mathtt{F}$. Such a tool would enable sudden climatic changes to be detected and quantified in real time, inspired by the methods used in finance to measure market volatility. The ability to anticipate and characterize these fluctuations would open up new opportunities for dynamic management of energy resources and improved adaptation strategies to extreme weather variations. Moreover, future work should aim to develop an approach that does not impose the constraint $ \mathtt{sCV} \to 0 \implies \left| \mathtt{I} - \mathtt{I}_{\mathtt{clr}} \right| \ll \left| \mathtt{I} - d \times \mathtt{I}_{\mathtt{clr}} \right| $. This requires to find a function $f(x)=\mathtt{sCV}$, with $x = \sqrt{\frac{1}{n} \sum_{i=1}^{n} (\mathtt{I}(t_i) - \mathtt{I}_{\mathtt{clr}}(t_i))^2}$ compatible with assumptions ($\gamma = \sqrt{\sigma^2_{\mathtt{clr}} + \mu_{\mathtt{clr}}^2} $): $ f(x) \to 0 $ as $ x \to 0^+ $, $ 0 < f(x) < 1 $ for $ x > 0 $, $ f(\frac{1}{\sqrt{2}} (d-1) \gamma) = 1 $, and $ f(x) \to 0 $ as $ x \to (\alpha - 1) \gamma $ for all $ \alpha \in [d,1] $.  This formulation could be significantly more complex, and it is essential to assess its impact through quantitative evaluation.

\section*{Acknowledgements}
This research was partially funded by the \texttt{ANR} under grant \texttt{SAPHIR} project (\texttt{ANR} reference: $\texttt{Sensor Augmented weather Prediction at HIgh-Resolution [ANR-21-CE04-0014]}$), whose support is gratefully acknowledged. The authors would also like to express their sincere thanks to \texttt{Elsevier} for providing access to the \texttt{Scopus} bibliographic database and for their valuable support and responsiveness during the integration and use of the \texttt{Scopus API}.

\appendix
\section{Derivation of $\sigma (\mathtt{k_c})$}
\label{perez}
For a raw time series $Y(t)$, the variance of consecutive differences, $\Delta Y = Y(t) - Y(t-1)$, is closely linked to the error metric $\mathtt{nRMSE}$. In the context of global horizontal irradiance $(\mathtt{I}$), and considering the clear-sky index $\mathtt{k_c}$ as a stationary process and defined as:
\begin{equation}
    \mathtt{k_c}(t) = \frac{\mathtt{I}(t)}{\mathtt{I}_{\mathtt{clr}}(t)},
\end{equation}
where $\mathtt{I}_{\mathtt{clr}}$ represents the theoretical clear-sky irradiance, the variance of $\Delta \mathtt{k_c} = \mathtt{k_c}(t) - \mathtt{k_c}(t-1)$ is expressed as:
\begin{equation}
    \sigma^2(\Delta \mathtt{k_c}) = 2\sigma^2(\mathtt{k_c})   (1 - \rho(1)),
\end{equation}
where $\rho(1)$ is the lag-1 autocorrelation of $\mathtt{k_c}$ and $\sigma^2(\mathtt{k_c})$ is the variance of $\mathtt{k_c}$. 
Assuming $\Delta \mathtt{k_c}$, and $\mathtt{\mathtt{I}_{\mathtt{clr}}(t)}$ are uncorrelated, the $\mathtt{RMSE}$ of the smart persistence model predicting $\mathtt{I}$ can be derived as:
\begin{equation}
    \mathtt{RMSE_{SP}} = \sqrt{2}   \sigma(\mathtt{k_c})   \sqrt{1 - \rho(1)}   \sqrt{\sigma^2(\mathtt{I}_{\mathtt{clr}}) + \mu_{\mathtt{clr}}^2}=\sigma(\Delta \mathtt{k_c}) \mu_{\mathtt{clr}}\sqrt{1+\frac{\sigma^2_{\mathtt{clr}}}{\mu^2_{\mathtt{clr}}}},
\end{equation}
where $\hat{\mathtt{I}}(t)$ is the predicted value at time $t$ given by $\hat{\mathtt{I}}(t) = \mathtt{k_c}(t-1)   \mathtt{I}_{\mathtt{clr}}(t).$ 
Similarly, the link between $\sigma(\Delta \mathtt{k_c})$ and $\mathtt{RMSE_{SP}}$ is $\mathtt{RMSE_{SP}} \propto \sigma(\Delta \mathtt{k_c})$, which highlights the role of temporal variability in both $\mathtt{k_c}$ and $\mathtt{I}_{\mathtt{clr}}$.

\section{Desirable Properties of Solar Irradiance Variability Estimators}
\label{properties2}
Robust estimators should meet the following criteria:
\begin{itemize}
    \setlength{\itemsep}{4pt} 
    \setlength{\parskip}{0pt}
    \setlength{\parsep}{0pt}
    \item[$\checkmark$] \cyr{The metric should be close to zero under clear-sky conditions, reflecting the relatively smooth and deterministic irradiance pattern in the absence of clouds. However, minor residual variability may arise from atmospheric factors such as aerosols not fully captured by clear-sky models};
    \item[$\checkmark$] It should approach 1 under extreme conditions, such as rapid alternation between clear-sky and overcast periods;
    \item[$\checkmark$] The estimator must avoid singularities or undefined behaviors, such as divisions by zero;
    \item[$\checkmark$] It must remain reliable despite minor inaccuracies in the estimation of clear-sky irradiance;
    \item[$\checkmark$] Overcast skies exhibit extremely low variability,  highlighting the need for metrics that accurately reflect such conditions;
    \item[$\checkmark$] It must exhibit a monotonic relationship with the ability to predict irradiation (directly or after adjustments, \textit{e.g.}, forecastability);
    \item[$\checkmark$] When variability increases, forecastability and predictability should decrease, unless the variability arises solely from deterministic and periodic behavior, in which case they remain equivalent;
    \item[$\checkmark$] \cyr{It should primarily depend on cloud occurrences (cloud-edge effects, transitions), unlike forecastability metrics which are also influenced by the time horizon or inclination};
    \item[$\checkmark$] The metric should be computationally efficient, allowing for real-time or near-real-time estimation in operational settings;
    \item[$\checkmark$] The metric should adapt to various temporal and spatial resolutions, ensuring consistent assessments across diverse applications.
\end{itemize}
These properties ensure that the estimator can support a range of applications, including forecasting, validation, and operational planning.

\section{Mathematical Properties of \texttt{sCV}}
\label{properties}
The stochastic coefficient of variation ($\mathtt{sCV}$) is designed to be a normalized measure of variability relative to the clear-sky irradiance ($\mathtt{I}_{\mathtt{clr}}$) and exhibits desirable properties for analyzing solar irradiance signals.
The $\mathtt{sCV}$ is defined as:
\begin{equation}\
\mathtt{sCV} =  \frac{1}{|d-1|}\frac{\sqrt{\frac{1}{n} \sum_{i=1}^n \left( \mathtt{I}(t_i) - \mathtt{I}_{\mathtt{clr}}(t_i) \right)^2}}{\mathrm{RMS}_{\mathtt{clr}}},
\end{equation}
where the ``cleared'' Root Mean Square is défined by $\mathrm{RMS}_{\mathtt{clr}} =\sqrt{\frac{\sigma_{\mathtt{clr}}^2 + \mu_{\mathtt{clr}}^2}{2}}$, with $\sigma_{\mathtt{clr}}$ and $\mu_{\mathtt{clr}}$ being the standard deviation and mean of $\mathtt{I}_{\mathtt{clr}}$, respectively.
The $\mathtt{sCV}$ is \textit{not} linear. While the variance of a linear combination of uncorrelated variables is:
\begin{equation}
\sigma^2(a \cdot \mathtt{I}_1 + b \cdot \mathtt{I}_2) = a^2 \cdot \sigma^2(\mathtt{I}_1) + b^2 \cdot \sigma^2(\mathtt{I}_2),
\end{equation}
normalizing by the $\mathrm{RMS}_{\mathtt{clr}}$ results in:
\begin{equation}
\mathtt{sCV}(a \cdot \mathtt{I}_1 + b \cdot \mathtt{I}_2) \neq |a| \cdot \mathtt{sCV}(\mathtt{I}_1) + |b| \cdot \mathtt{sCV}(\mathtt{I}_2).
\end{equation}
Similarly, the additivity of variances:
\begin{equation}
\sigma^2(\mathtt{I}_1 + \mathtt{I}_2) = \sigma^2(\mathtt{I}_1) + \sigma^2(\mathtt{I}_2),
\end{equation}
does not translate into a simple additivity property for $\mathtt{sCV}$ due to the normalization by $\mathrm{RMS}_{\mathtt{clr}}$. As such, $\mathtt{sCV}$ of the sum is not the sum of the $\mathtt{sCV}$s.
The $\mathtt{sCV}$ is bounded between $[0, 1]$. Its normalization ensures consistency across varying conditions, $\mathtt{sCV} = 0$ corresponds to no variability relative to the reference and $\mathtt{sCV} = 1$ occurs in scenarios with maximal variability (\textit{e.g.}, alternating between $\mathtt{I}_{\mathtt{clr}}$ and $d \cdot \mathtt{I}_{\mathtt{clr}}$).
The lack of linearity and additivity makes $\mathtt{sCV}$ less suitable for hierarchical analyses or scenarios involving aggregation (\textit{e.g.} multiple-site variability). However, it remains a valid and robust tool for comparing variability relative to a dynamic reference, making it particularly relevant for site-specific analyses or intra-day variability studies.

\section{From Variability to Volatility}
\label{volatility}
The stochastic coefficient of variation ($\mathtt{sCV}$) shares conceptual similarities with the concept of volatility in financial time series \cite{10.2307/2298006}, where variability is dynamically evaluated over rolling windows to capture short-term fluctuations. In the context of solar irradiance, $\mathtt{sCV}$ can similarly be computed over a sliding window $[t - \Delta t, t]$ of length $\Delta t$, allowing it to adapt to temporal changes in irradiance patterns.
Within each window, the variance relative to the clear-sky reference $\mathtt{I}_{\mathtt{clr}}(t)$ is defined as:
\begin{equation}
\sigma^2_{\Delta t}(\mathtt{I}) = \frac{1}{\Delta t} \int_{t-\Delta t}^t \left(\mathtt{I}(\tau) - \mathtt{I}_{\mathtt{clr}}(\tau)\right)^2 \, d\tau,
\end{equation}
with the corresponding standard deviation given by $\sigma_{\Delta t}(\mathtt{I}) = \sqrt{\sigma^2_{\Delta t}(\mathtt{I})}.$
The localized stochastic coefficient of variation is then expressed as $ \mathtt{sCV}_{\Delta t} = \alpha \cdot \sigma_{\Delta t}(\mathtt{I}),$
where $\alpha$ is a scaling factor ensuring consistency with the global $\mathtt{sCV}$ definition.
This sliding window computation enables $\mathtt{sCV}$ to capture a localized version of variability relative to the clear-sky trend. As $\Delta t \to \infty$, this measure converges to the global definition of $\mathtt{sCV}$ under stationarity assumptions \cite{GORDON198935}. For finite $\Delta t$, it provides a real-time assessment of instantaneous variability, making it particularly suitable for dynamic or non-stationary contexts.
By analogy with volatility in financial markets, $\mathtt{sCV}$ offers energy system operators a robust tool to dynamically monitor and adapt to fluctuations in solar resources. Its capacity to reflect both short-term and long-term variability enhances its applicability in real-time operations, forecasting, and grid stability planning.

\bibliographystyle{elsarticle-num} 
\bibliography{Bib}

\end{document}